\documentclass[12pt]{article}
\usepackage{epsfig}
\usepackage{axodraw}
\def\beq{\begin{equation}}
\def\eeq{\end{equation}}
\def\bea{\begin{eqnarray}}
\def\eea{\end{eqnarray}}
\def\bq{\begin{quote}}
\def\eq{\end{quote}}

\def\bq{\begin{quote}}
\def\eq{\end{quote}}

\def\ma{m_{\chi_a}}
\def\mb{m_{\chi_b}}

\def\etal{{\it et al.}}

\def\bq{\begin{quote}}
\def\eq{\end{quote}}

\def\gappeq{\mathrel{\rlap {\raise.5ex\hbox{$>$}}
{\lower.5ex\hbox{$\sim$}}}}

\def\lappeq{\mathrel{\rlap{\raise.5ex\hbox{$<$}}
{\lower.5ex\hbox{$\sim$}}}}

\def\bbz{fa Z \kern-8.9pt Z}

\begin{document}

\baselineskip 18pt
\newcommand{\sheptitle}
{
On the Large Higher Order Corrections to Polarised Chargino Production
}

\newcommand{\shepauthor}
{Marco A. D\'\i az$^{1}$ and Douglas A. Ross${^2}$}

\newcommand{\shepaddress}
{${^1}$Departmento de F\'\i sica, Universidad Cat\'olica de Chile,
Av. Vicu\~na Mackenna 4860, Santiago, Chile\\
$^{2}$Department of Physics
and Astronomy, University of Southampton, Southampton, SO17 1BJ,
U.K.}

\newcommand{\shepabstract}
{We calculate the radiative corrections to helicity amplitudes for
chargino production in electron-positron collisions. We include all weak
self-energy, triangle, and box diagrams, and find that the three are 
important and should be included. We present results in the form of
differential cross-sections for four supersymmetry benchmark models,
and find usually large corrections and in some cases huge corrections.
We conclude that in order to extract the underlying parameters of the 
chargino sector from collider data taken at a future Linear Collider,
a complete theoretical one--loop production cross-section should be used.
}

\begin{titlepage}
\begin{flushright}
hep-ph/0205257\\
\end{flushright}
\begin{center}
{\large{\bf \sheptitle}}
\bigskip \\ \shepauthor \\ \mbox{} \\ {\it \shepaddress} \\ \vspace{.5in}
{\bf Abstract} \bigskip \end{center} \setcounter{page}{0}
\shepabstract
\end{titlepage}

\section{Introduction}   \label{intro}

Experimental Particle Physics in the last decade has clarified the basic 
structure of the strong, weak, and electromagnetic interactions. Gauge 
symmetry based on the group $SU(3)\times SU(2)\times U(1)$ have been 
successfully tested against the theoretical model now called the Standard 
Model (SM), with the only exception of the symmetry breaking mechanism. 
Precision measurements on the different observables provided important 
information on the top quark mass even before its discovery. In the same 
way, information about the Higgs mass and unification of gauge couplings 
is being extracted from these measurements. This could not be done if 
experimental precision measurements were not accompanied by precise 
theoretical calculations, which include one-loop corrections to the 
different observables. This success of the SM is not expected to hold at
higher energies, and new physics is required for the model to be consistent.
Precision measurements studies indicate that the threshold for this new 
physics is not far from the reach of FERMILAB, or at least of the CERN LHC.

The most studied extension of the SM is the Minimal Supersymmetric Standard
Model (MSSM), which includes a symmetry between bosons and fermions. 
Charginos are mixings of the supersymmetric partners of the charged Higgs 
and gauge bosons and are expected to be amongst the lightest superpartners. 
Their masses are determined by the supersymmetric Higgs mass parameter $\mu$,
the ratio between the Higgs vacuum expectation values $\tan\beta$, and 
the gaugino mass $M$, which is one of the soft supersymmetry breaking mass 
parameters introduced to break supersymmetry. If charginos are discovered,
their masses and couplings will only be measured with precision at an
electron--positron collider. The precision measurements that a future 
Linear Collider \cite{LC} is capable of, will give information on the 
underlying theory, and a determination of the fundamental parameters will 
be achieved only if precise theoretical calculation are performed to match 
the experimental precision.

Radiative corrections to chargino observables started with the one--loop
calculation of chargino masses in \cite{PP}, which were complemented later 
with similar studies in \cite{mass1loop}, and with phenomenological 
consequences in the analysis of LEP data in \cite{EFGOS}. Chargino mass 
determination based on an analysis using background cut indicates a 1\% 
precision or less \cite{mcuts}, while threshold scans can potentially give
0.1\% precision on the chargino mass determination \cite{threshold}. 
Clearly, one-loop corrections to chargino masses must be included, since 
they are typically of several percent.

Quantum corrections to chargino production in electron--positron annihilation 
are more recent. Corrections to the unpolarized cross-section due to quarks
and squarks in the loops were calculated in ref.~\cite{DKR,KNPY} and proved 
to be important. Radiative corrections to chargino pair production with 
polarized electron and positron beams were calculated for the first time in
\cite{DKR2} including only quarks and squarks in the loops, and large 
corrections were found especially for right handed electrons. A complete
one-loop corrected chargino pair production cross-section with polarized 
beams, but still without considering the chargino helicities, was reported
in ref.~\cite{BH}, where large corrections were found and the importance 
of box diagrams was stressed. Radiative corrections to chargino decays have 
also been calculated \cite{decay}.

The importance of the chargino (and neutralino) helicities and the spin
correlation between production and decay has been pointed out extensively
\cite{gudrid}. Studies on how to extract the fundamental parameters of the 
chargino sector from precision measurements on chargino masses and 
production cross-sections are reported in \cite{kal,chaparams}. These 
studies were done at tree level. A first step towards the complete 
one-loop corrections to chargino production helicity amplitudes was done 
in ref.~\cite{prototype}, where expressions for all diagrams, including 
triangles and boxes, were given in terms of Veltman-Passarino functions as 
prototype diagrams. In addition, a general formula was given for the 
helicity amplitudes valid for chargino production of equal or different 
masses. In this paper we present the first calculation for one--loop
corrections to chargino production helicity amplitudes with polarized beams,
where we include all weak contributions in the form of self energies, 
triangles and box diagrams.
We have organized the calculation in terms of a set of prototype graphs
described in detail in  ref.~\cite{prototype}. The various sets of internal
particles that can contribute to each of these prototypes is given in
the Appendix. \footnote{A FORTRAN package which calculates the contributions
from each of these prototypes can be found at: \\
http://www.hep.phys.soton.ac.uk/hepwww/staff/D.Ross/chipackage/chipackage.html
.}

In this paper, we have restricted the calculation to the corrections due
to weak interactions and loops of super-partners. Pure QED corrections
involving loops of photons have been omitted.
Such (virtual) QED corrections will introduce infrared divergences which
will cancel when the Bremsstrahlung process is taken into account.
The remnant QED correction is then sensitive to the energy resolution
of the final state charginos. Moreover, the QED corrections will
contain enhanced logarithmic terms of order $\alpha_{em}/\pi \ln(s/m_e^2)$
arising from initial state radiation (ISR) in which the photon is emitted
parallel to one of the incoming leptons. Such large corrections are 
universal to {\it all} electron-proton annihilation processes and 
(as in ref.\cite{BH}) we assume that these will have been accounted for
at the data analysis level. Having done this the remaining QED corrections
are expected to be genuinely of order $\alpha_{em}/\pi$ provided the
final state energy resolution is not taken too small.

A delicate matter  was raised in ref.\cite{BH} concerning the
UV finiteness of the calculation in which part of the corrections are omitted.
For on-shell renormalization, such as that used in ref.\cite{BH}, 
in which the counterterms required to cancel any UV divergence are obtained
from the higher order corrections to some physical process in which
the external particles are on-shell, there would indeed be a remnant
UV divergence should any part of the higher order corrections be omitted.
In our case, where we use the $\overline{DR}$ scheme
to define all renormalized couplings (and hence all SUSY parameters),
all the UV divergences are subtracted by simply removing the pole
 part (plus $\ln(4\pi)-\gamma_E$) from any divergent integral. 
The effect of omitting photon corrections whilst implicitly including
photino corrections (through the neutralino exchanges) is that the counterterms
violate supersymmetry invariance, e.g. there will be a different counterterm
associated with the $Z-W-W$ vertex from that associated with the
 $Z-\tilde{W}-\tilde{W}$ vertex (where $\tilde{W}$ is the wino component
of the charginos). The breaking of this symmetry in the
renormalization programme is no worse than the breaking of the gauge symmetry
which occurs in the on-shell scheme in which the counterterm associated with 
the $Z-\tilde{\chi}-\tilde{\chi}$ vertex is different from that of
 the $Z-e-e$ vertex. In other words, we omit a contribution to
the loop corrections and to the corresponding counterterm in such a way
that the finite difference is genuinely of order  $\alpha_{em}/\pi$.
Whereas the use of the  $\overline{DR}$ renormalization scheme
has the disadvantage of not relating the renormalized couplings
directly to a physically measurable quantity, it has the advantage 
of prescribing how these parameters may be deduced from any
given relevant measurement.
Once any measurement has been performed, from which a (combination of)
coupling can be deduced, it is a simple matter to derive a dictionary 
between the value deduced form the experiment and its value in the
 $\overline{DR}$ scheme. In the absence of any such experimental data,
 however, nothing is gained by relating  the parameters 
 in the $\overline{DR}$ scheme to their values derived from any one
particular set of experimental measurements. The quoted values of all 
parameters in this paper are therefore to be understood to be in the
 $\overline{DR}$.

\section{The Helicity Amplitudes}

Consider the production of a pair of charginos in electron--positron 
annihilation:
\begin{equation}
e^+(p_2)\quad e^-(p_1) \quad \longrightarrow \quad 
\tilde\chi^+_b(k_2) \quad \tilde\chi^-_a(k_1)
\end{equation}
The electron has momentum $p_1$ and polarization $\alpha=L,R$, while the
positron has momentum $p_2$ and opposite polarization. The (positively
charged) chargino has mass $\mb$, momentum $k_2$ and helicity $\lambda_2$,
while the (negatively charged) anti-chargino has a mass $\ma$, momentum
$k_1$, and helicity $\lambda_1$. With this notation, the scattering 
amplitude is written as
\beq {\cal A}^\alpha_{\lambda_2,\lambda_1} \ = \ 
   \frac{2}{s} L^\mu_{\alpha}\, 
    Q_{i\,\mu}^{\alpha} \, \langle k_2, \lambda_2 | \Gamma^i
 | k_1, \lambda_1 \rangle . \label{me1} \eeq
where $\sqrt{s}$ is the center of mass energy.
These amplitudes are normalized as in \cite{kal}, such that the differential
cross-section is given by
\beq \frac{d\sigma(\alpha,\lambda_2,\lambda_1)}{d\cos\theta}
  \ = \ \frac{\lambda^{1/2}(s,\ma^2,\mb^2)}{128 \, \pi \, s} 
 \left| {\cal A}^\alpha_{\lambda_2,\lambda_1} \right|^2, \eeq
where
$$ \lambda(x,y,z) \ \equiv \ x^2  +  y^2  +  z^2 
  -  2  x  y     -  2  x  z  -  2  y  z $$
and $\theta$ is the scattering angle between the electron and the chargino
momenta.

As was shown in ref.~\cite{prototype}, the contribution from any Feynman 
graph to such an amplitude can always be expressed in this form by making 
a suitable Fierz transformation where necessary. Here $ L^\mu_{R(L)}$ is 
the leptonic matrix element
$$  L^\mu_{R(L)} \ = \ \bar{v}(p_2) \gamma^\mu \frac{(1\pm \gamma^5)}{2}
    u(p_1). $$
Since the leptons are considered to be massless these two are the only
possible structures for the lepton factor. Following \cite{prototype}, 
the chargino factor is written as the sum of matrix elements of five 
possible $\gamma-$matrix structures $\Gamma_i, \ i=1 \cdots 5$ are given by
\begin{eqnarray} \Gamma^1 & = & \frac{(1+\gamma^5)}{2} \nonumber \\
 \Gamma^2 & = & \frac{(1-\gamma^5)}{2}  \nonumber \\
 \Gamma^3 & = & \gamma^\nu\frac{(1+\gamma^5)}{2}  \nonumber \\
 \Gamma^4 & = & \gamma^\nu\frac{(1-\gamma^5)}{2}  \nonumber \\
 \Gamma^5 & = & -i \, \sigma^{\nu\rho} \end{eqnarray}
The coefficients of these structures, $ Q_{i\,\mu}^{\alpha}$, are tensors 
which can be reduced to the following structures, in terms of scalar 
quantities ${\cal Q}^{\alpha}_{i\,j}, \ i=1...5,\ j=1,2, \ \alpha=L,R$,  as 
follows
\begin{eqnarray}
Q^{\alpha\mu}_1 & = & {\mathcal Q}^{\alpha}_1 \, k_-^\mu \ 
  \nonumber \\
Q^{\alpha\mu}_2 & = & {\mathcal Q}^{\alpha}_2 \, k_-^\mu \ 
    \nonumber \\
Q^{\alpha\mu\nu}_3 & = & {\mathcal Q}^{\alpha}_{31} \, g^{\mu\nu} \  \, + 
             {\mathcal Q}^{\alpha}_{32} \, k_-^\mu \,  p^\nu \nonumber \\
Q^{\alpha\mu\nu}_4 & = & {\mathcal Q}^{\alpha}_{41} \, g^{\mu\nu} \  \, + \ 
             {\mathcal Q}^{\alpha}_{42} \, k_-^\mu \, p^\nu \nonumber \\
Q^{\alpha\mu\nu\rho}_5 
& = & {\mathcal Q}^{\alpha}_{51} \, g^{\mu\nu} \, p^\rho \  \, -  \ 
   i \,  {\mathcal Q}^{\alpha}_{52} \, \epsilon^{\mu\nu\rho\tau} \, p_\tau,
   \end{eqnarray}
where $ k_-^\mu=(k_1^\mu-k_2^\mu)$ and $ p^\mu=(p_1^\mu-p_2^\mu)$. Any 
other structure can be expressed in terms of the above quantities, by 
exploiting the fact that the leptonic current is conserved and that the 
matrix elements of  $\Gamma^i$ are taken between on-shell chargino states. 
These Q-charges ${\cal Q}^{\alpha}_{i\,j}$ are higher order generalizations 
of the Q-charges $Q_{LL}$, $Q_{LR}$, $Q_{RL}$, and $Q_{RR}$ defined for 
example in ref.~\cite{kal} for the tree-level case.
In our notation, at the tree level only ${\mathcal Q}^{\alpha}_{31}$ and 
${\mathcal Q}^{\alpha}_{41}$ are non-zero. Furthermore 
${\mathcal Q}^{\alpha}_{32}, \ {\mathcal Q}^{\alpha}_{42}, \ 
{\mathcal Q}^{\alpha}_{51}$ and ${\mathcal Q}^{\alpha}_{52}$ do not occur 
in self-energy or vertex correction graphs, but only arise when boxes are 
taken into consideration.

Helicity amplitudes are given for the general case in ref.~\cite{prototype}.
Here we list them for the case $m_{\chi_a}=m_{\chi_b}$. Helicity amplitudes
are denoted by ${\cal A}^{\alpha}_{\lambda_2\lambda_1}$, where $\alpha=L,R$
is the polarization of the electron, and $\lambda_2\lambda_1$ are the 
helicities of the chargino and anti-chargino respectively. For left handed 
electrons we have:
\begin{eqnarray}
{\cal A}^L_{++}&=&
-{\cal Q}^L_1 \sqrt{s}\,v\,(1-v) \sin\theta
+{\cal Q}^L_2 \sqrt{s}\,v\,(1+v) \sin\theta 
\nonumber\\ &&
+({\cal Q}^L_{31}+{\cal Q}^L_{41})\sqrt{1 - v^2}\, \sin\theta
-({\cal Q}^L_{32}+{\cal Q}^L_{42})\,s\,v\,\sqrt{1 - v^2}\,
\sin\theta\cos\theta 
\nonumber\\ &&
-2{\cal Q}^L_{51}\sqrt{s}\,v\,\sin\theta
+4{\cal Q}^L_{52}\sqrt{s}\,\sin\theta
\label{aLpp}
\\ & & \nonumber \\
{\cal A}^L_{+-}&=&
-{\cal Q}^L_{31}(1+v)\,(1+\cos\theta)
-{\cal Q}^L_{32}\,s\,v\,(1+v)\,\sin^2\theta
\nonumber\\ &&
-{\cal Q}^L_{41}(1-v)\,(1+\cos\theta)
-{\cal Q}^L_{42}\,s\,v\,(1-v)\,\sin^2\theta
\nonumber\\ &&
-4{\cal Q}^L_{52}\sqrt{s}\,\sqrt{1 - v^2}\,(1+\cos\theta)
\label{aLpm}
\\ & & \nonumber \\
{\cal A}^L_{-+}&=&
+{\cal Q}^L_{31}(1-v)\,(1-\cos\theta)
-{\cal Q}^L_{32}\,s\,v\,(1-v)\,\sin^2\theta
\nonumber\\ &&
+{\cal Q}^L_{41}(1+v)\,(1-\cos\theta)
-{\cal Q}^L_{42}\,s\,v\,(1+v)\,\sin^2\theta
\nonumber\\ &&
+4{\cal Q}^L_{52}\sqrt{s}\,\sqrt{1 - v^2}\,(1-\cos\theta)
\label{aLmp}
\\ & & \nonumber \\
{\cal A}^L_{--}&=&
-{\cal Q}^L_1 \sqrt{s}\,v\,(1+v) \sin\theta
+{\cal Q}^L_2 \sqrt{s}\,v\,(1-v) \sin\theta 
\nonumber\\ &&
-({\cal Q}^L_{31}+{\cal Q}^L_{41})\sqrt{1 - v^2}\, \sin\theta
+({\cal Q}^L_{32}+{\cal Q}^L_{42})\,s\,v\,\sqrt{1 - v^2}\,
\sin\theta\cos\theta 
\nonumber\\ &&
-2{\cal Q}^L_{51}\sqrt{s}\,v\,\sin\theta
-4{\cal Q}^L_{52}\sqrt{s}\,\sin\theta
\label{aLmm}
\end{eqnarray}
where $v$ is the chargino velocity given by
\begin{equation}
v=\sqrt{1-{{4m_\chi^2}\over s}}
\end{equation}
The helicity amplitudes for right handed electrons are:
\begin{eqnarray}
{\cal A}^R_{++}&=&
-{\cal Q}^R_1 \sqrt{s}\,v\,(1-v) \sin\theta
+{\cal Q}^R_2 \sqrt{s}\,v\,(1+v) \sin\theta 
\nonumber\\ &&
+({\cal Q}^R_{31}+{\cal Q}^R_{41})\sqrt{1 - v^2}\, \sin\theta
-({\cal Q}^R_{32}+{\cal Q}^R_{42})\,s\,v\,\sqrt{1 - v^2}\,
\sin\theta\cos\theta 
\nonumber\\ &&
+2{\cal Q}^R_{51}\sqrt{s}\,v\,\sin\theta
-4{\cal Q}^R_{52}\sqrt{s}\,\sin\theta
\label{aRpp}
\\ & & \nonumber \\
{\cal A}^R_{+-}&=&
+{\cal Q}^R_{31}(1+v)\,(1-\cos\theta)
-{\cal Q}^R_{32}\,s\,v\,(1+v)\,\sin^2\theta
\nonumber\\ &&
+{\cal Q}^R_{41}(1-v)\,(1-\cos\theta)
-{\cal Q}^R_{42}\,s\,v\,(1-v)\,\sin^2\theta
\nonumber\\ &&
-4{\cal Q}^R_{52}\sqrt{s}\,\sqrt{1 - v^2}\,(1-\cos\theta)
\label{aRpm}
\\ & & \nonumber \\
{\cal A}^R_{-+}&=&
-{\cal Q}^R_{31}(1-v)\,(1+\cos\theta)
-{\cal Q}^R_{32}\,s\,v\,(1-v)\,\sin^2\theta
\nonumber\\ &&
-{\cal Q}^R_{41}(1+v)\,(1+\cos\theta)
-{\cal Q}^R_{42}\,s\,v\,(1+v)\,\sin^2\theta
\nonumber\\ &&
+4{\cal Q}^R_{52}\sqrt{s}\,\sqrt{1 - v^2}\,(1+\cos\theta)
\label{aRmp}
\\ & & \nonumber \\
{\cal A}^R_{--}&=&
-{\cal Q}^R_1 \sqrt{s}\,v\,(1+v) \sin\theta
+{\cal Q}^R_2 \sqrt{s}\,v\,(1-v) \sin\theta 
\nonumber\\ &&
-({\cal Q}^R_{31}+{\cal Q}^R_{41})\sqrt{1 - v^2}\, \sin\theta
+({\cal Q}^R_{32}+{\cal Q}^R_{42})\,s\,v\,\sqrt{1 - v^2}\,
\sin\theta\cos\theta 
\nonumber\\ &&
+2{\cal Q}^R_{51}\sqrt{s}\,v\,\sin\theta
+4{\cal Q}^R_{52}\sqrt{s}\,\sin\theta
\label{aRmm}
\end{eqnarray}
At tree level, these expressions reduce to
\begin{eqnarray}
{\cal A}^{L,0}_{++}&=&
({\cal Q}^{L,0}_{31}+{\cal Q}^{L,0}_{41})\sqrt{1 - v^2}\, \sin\theta
\nonumber\\
{\cal A}^{L,0}_{+-}&=&
-[{\cal Q}^{L,0}_{31}(1+v)+{\cal Q}^{L,0}_{41}(1-v)]\,(1+\cos\theta)
\nonumber\\
{\cal A}^{L,0}_{-+}&=&
[{\cal Q}^{L,0}_{31}(1-v)+{\cal Q}^{L,0}_{41}(1+v)]\,(1-\cos\theta)
\nonumber\\
{\cal A}^{L,0}_{--}&=&
-({\cal Q}^{L,0}_{31}+{\cal Q}^{L,0}_{41})\sqrt{1 - v^2}\, \sin\theta
\end{eqnarray}
for the left handed electron amplitudes, and
\begin{eqnarray}
{\cal A}^{R,0}_{++}&=&
({\cal Q}^{R,0}_{31}+{\cal Q}^{R,0}_{41})\sqrt{1 - v^2}\, \sin\theta
\nonumber\\
{\cal A}^{R,0}_{+-}&=&
[{\cal Q}^{R,0}_{31}(1+v)+{\cal Q}^{R,0}_{41}(1-v)]\,(1-\cos\theta)
\nonumber\\
{\cal A}^{R,0}_{-+}&=&
-[{\cal Q}^{R,0}_{31}(1-v)+{\cal Q}^{R,0}_{41}(1+v)]\,(1+\cos\theta)
\nonumber\\
{\cal A}^{R,0}_{--}&=&
-({\cal Q}^{R,0}_{31}+{\cal Q}^{R,0}_{41})\sqrt{1 - v^2}\, \sin\theta
\end{eqnarray}
for the right handed electrons.

This coincides with the expressions given
in ref.~\cite{kal} (after allowing for the fact that in 
ref.~\cite{kal} the negatively charged chargino is taken to be the particle
and the positively charged one the antiparticle, whereas our convention
is {\it vice versa}.)



\section{Renormalization Procedure}

We regularize divergent diagrams using dimensional reduction $\overline{DR}$.
In each graph, divergences are contained in the parameter
\begin{equation}
\Delta={2\over{4-n}}+\ln 4\pi-\gamma_E
\end{equation}
where $n$ is the number of space-time dimensions and $\gamma_E$ is the 
Euler's constant.
The renormalization subtraction point is taken to be $\mu^2=M_Z^2$.


As in \cite{DKR}, we organize the self-energy and triangle contributions
in form factors for the $Z$--chargino--chargino vertex, given by
\begin{equation}
{\cal G}_{Z\chi\chi}=
F_{Z0}^+ \,\gamma^\mu \frac{(1+\gamma^5)}{2}
\, + \,   F_{Z0}^- \,\gamma^\mu \frac{(1-\gamma^5)}{2}
\, + \, F_{Zk}^+ \,k_-^\mu \frac{(1+\gamma^5)}{2}
\, +  \, F_{Zk}^- \,k_-^\mu \frac{(1-\gamma^5)}{2}
\end{equation}
and similarly for the photon--chargino--chargino vertex, and form factors 
in the $e^{\pm}$--sneutrino--chargino vertices:
\begin{equation}
{\cal G}_{\tilde\nu e\chi}^+=
F_{\tilde{\nu}}^+ \frac{(1+\gamma^5)}{2} C \qquad
{\cal G}_{\tilde\nu e\chi}^-=
F_{\tilde{\nu}}^- C^{-1} \frac{(1-\gamma^5)}{2}
\end{equation}
where $C$ is the charged conjugation matrix.

%

The photon self-energy vanishes at zero momentum
by virtue of gauge invariance so that after subtraction
in the  $\overline{DR}$ scheme  contributes to the photon form factor 
according to

\begin{equation}
\Delta F_{\gamma 0}^{\pm}=
e{{A_{\gamma\gamma}(s)}\over{s}}\delta_{ab}
\end{equation}
where $a$ and $b$ refer to the two species of charginos produced. 
The 
photon-$Z$ mixing is also subtracted in the  $\overline{DR}$ scheme
and contributes to the $Z$ form factor
\begin{equation}
\Delta F_{Z 0}^{\pm}=
e{{A_{Z\gamma}(s)}\over{s}}\delta_{ab}
\end{equation}
and to photon form factors:
\begin{equation}
\Delta F_{\gamma 0}^{\pm}=
-{g\over{c_W}}O'^{R(L)}_{ab}{{A_{Z\gamma}(s)}\over{s-m_Z^2}}
\end{equation}

The $Z$--boson self-energy is regularized with a subtraction at
$s=m_Z^2$:
\begin{equation}
\Delta F_{Z 0}^{\pm}=
-{g\over{c_W}}O'^{R(L)}_{ab}{{A_{ZZ}(s)-
A_{ZZ}(m_Z^2)}\over{s-m_Z^2}}
\end{equation}
and similarly for the sneutrino self-energy
\begin{equation}
\Delta F_{\tilde\nu}^{\pm}=
-{1\over2}gV_{b(a)1}{{A_{\tilde\nu\tilde\nu}(s)-
A_{\tilde\nu\tilde\nu}(m_{\tilde\nu}^2)}\over{t-m_{\tilde\nu}^2}}
\end{equation}
This guarantees that the parameters $m_Z$ and $m_{\tilde\nu}$
respectively refer to the physical (pole-)masses.

Chargino self-energy and mixing contribute to form factors in a more 
complicated way. 
Since these are external particles we have insisted that
the subtractions are performed on-shell, so that the renormalized
chargino fields are indeed physical fields.
 Details can be found in ref.~\cite{prototype,DKR}.


Ultraviolet divergences that occur in a few of the triangle graphs
 are subtracted in the $\overline{DR}$ scheme with subtraction point
$m_z$. Therefore, apart for the masses which are taken to be physical
and the weak-mixing angle whose renormalization is described above, all
other parameters are to be considered to be in the
 $\overline{DR}$ at $m_Z$.

We point out here that all couplings are now taken to be
couplings in the  $\overline{DR}$ scheme at the scale $\mu=M_Z$
{\it in the MSSM theory}. This means that the translation
of the values used here to those directly extracted
from experiment, such as neutral current 
neutrino scattering cross-sections
or the measured fine-structure constant will be slightly different from
that of the Standard Model (without the supersymmetric partners). 
For example, the treatment of the photon-$Z$ 
propagator system,
 described above,
guarantees that the propagators only have poles at zero and  $M_Z$,
but there is still some remnant of photon-$Z$ mixing at these poles.
We have checked numerically that the effect of a further subtraction
of the photon-$Z$ mixing propagator to remove this mixing has a negligible
numerical effect on our results.
Furthermore, the input SUSY parameters chosen
are assumed also to be
the corresponding values renormalized in this scheme at the same scale.
We expect the sensitivity to (reasonable) changes in renormalization scheme
to be genuinely of order $\alpha_W/\pi$ and to have no significant 
effect on our numerical results.

\section{CP Invariance}

Provided that the couplings are all taken to be real, the scattering 
amplitudes must be CP invariant. A consequence of this is that (for like 
species of produced charginos)
\begin{equation} 
{\cal A}_{++} \ =  \ \eta \,  {\cal A}_{--}, 
\end{equation}
where $\eta$ is a phase that depends on the phase convention taken for
the chargino spinors. In our case we have $\eta=-1$.

For this relation to hold at all scattering angles and all energies, we can 
see from eqs.~(\ref{aLpp}-\ref{aRmm}) that we require
\begin{eqnarray} 
{\cal Q}^{\alpha}_1 = {\cal Q}^{\alpha}_2\,, \qquad
{\cal Q}^{\alpha}_{51} = 0 \,,\qquad \alpha=L,R.
\label{CP1}
\end{eqnarray}
However, such remarkable cancellations do not occur on a graph-by-graph 
basis. Nevertheless, a set of graphs containing the same internal particles
must satisfy these relations by themselves, since a small
change in the input SUSY parameters would spoil any mutual cancellation
between different sets of graphs. This cancellation provides a highly
 non-trivial check both of the analytic expressions for the prototype
graphs, as well as the numerical implementation of the
Veltman-Passarino functions.
\bigskip

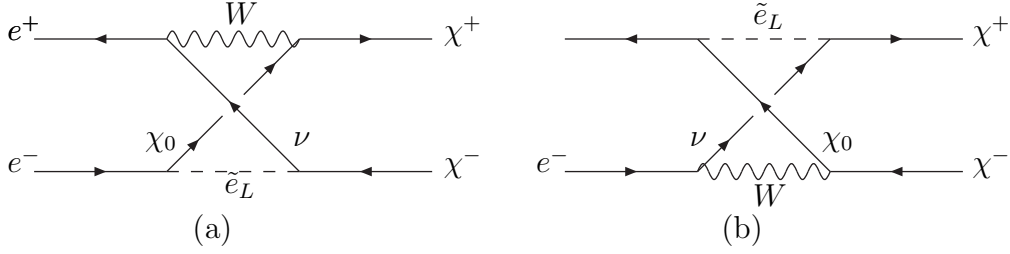
\begin{figure}[h]
\begin{picture}(400,80)
\ArrowLine(10,25)(60,25) \put(0,25){$e^-$}
\ArrowLine(60,25)(81,46) \ArrowLine(89,54)(110,75)   \put(52,35){$\chi_0$}
\ArrowLine(110,75)(160,75) \put(165,75){$\chi^+$}

\ArrowLine(160,25)(110,25)  \put(165,25){$\chi^-$}
\ArrowLine(110,25)(60,75)     \put(108,35){$\nu$}
\ArrowLine(60,75)(10,75) \put(0,75){$e^+$}

\Photon(60,75)(110,75){3}{6} \put(82,81){$W$}
\DashLine(60,25)(110,25){5}  \put(82,17){$\tilde{e}_L$}
\put(70,0){(a)}

\ArrowLine(210,25)(260,25) \put(200,25){$e^-$}
\ArrowLine(260,25)(281,46) \ArrowLine(289,54)(310,75)   \put(258,35){$\nu$}
\ArrowLine(310,75)(360,75) \put(365,75){$\chi^+$}

\ArrowLine(360,25)(310,25)  \put(365,25){$\chi^-$}
\ArrowLine(310,25)(260,75)     \put(308,35){$\chi_0$}
\ArrowLine(260,75)(210,75) \put(0,75){$e^+$}

\Photon(260,25)(310,25){3}{6} \put(282,12){$W$}
\DashLine(260,75)(310,75){5}  \put(282,81){$\tilde{e}_L$}
\put(270,0){(b)}

\end{picture}
\caption{\it Set of box diagrams with $W$, $\nu_e$, $\tilde e_L$, and 
$\chi^0_i$ inside the loop. A non trivial cancellation must occur in order 
to get CP invariance.} 
\label{CPfig1} \end{figure}
\begin{figure}[h]
\begin{picture}(400,80)
\ArrowLine(160,25)(110,25) \put(165,25){$e^-$} 
\ArrowLine(110,25)(60,25) \put(90,13){$\chi^0$}
\ArrowLine(60,25)(10,25)  \put(0,25) {$\chi^+$}

\ArrowLine(160,75)(110,75) \put(165,75){$\chi^-$} 
\ArrowLine(110,75)(60,75) \put(90,65){$\nu$}
\ArrowLine(60,75)(10,75)  \put(0,75) {$e^+$}

\Photon(60,25)(60,75){3}{6} \put(42,45){$W$}
\DashLine(110,25)(110,75){5} \put(113,45){$\tilde{e}_L$}

\put(70,30){$\leftarrow  l$}
\put(80,0){(a)}

\ArrowLine(360,25)(310,25) \put(365,25){$e^-$} 
\ArrowLine(310,25)(260,25) \put(290,13){$\nu$}
\ArrowLine(260,25)(210,25)  \put(200,25) {$\chi^+$}

\ArrowLine(360,75)(310,75) \put(365,75){$\chi^-$} 
\ArrowLine(310,75)(260,75) \put(290,65){$\chi^0$}
\ArrowLine(260,75)(210,75)  \put(200,75) {$e^+$}

\Photon(310,25)(310,75){3}{6} \put(315,45){$W$}
\DashLine(260,25)(260,75){5} \put(242,45){$\tilde{e}_L$}

\put(270,30){$\leftarrow  l$}
\put(280,0){(b)}

\end{picture}
\caption{\it Uncrossed diagrams corresponding to the boxes in 
Fig.~\ref{CPfig1}.} 
\label{CPfig2} \end{figure}
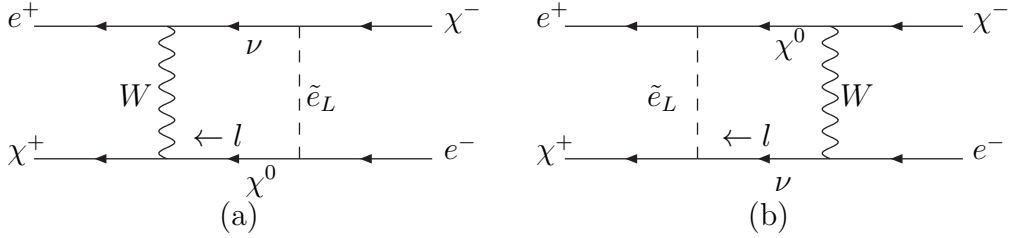

%
We demonstrate this with an example of a pair of box diagrams shown in
Fig. \ref{CPfig1}, which contain a $W$, a neutralino, a neutrino, and
a left-handed selectron inside the loop. Performing a crossing in the
$s$-channel these two graphs become prototypes 5a and 5b respectively.
A further crossing in the $t$-channel reduces them to the uncrossed
box graphs shown in Fig.\ref{CPfig2}. In these figures we have indicated
the rooting of the loop momentum, $l$, that we have taken.
\bigskip

%
Concentrating only on the coefficients ${\cal Q}^L_1, \ 
{\cal Q}^L_2$
and ${\cal Q}^L_{51}$, we obtain contributions from  Fig.\ref{CPfig2}(a)
in terms of the Veltman-Passarino functions, $D0, \ D1, \ D2$, with zero,
one, and two powers of loop momentum in the numerators respectively
(defined in detail in \cite{prototype}).
\footnote{Unfortunately there is a misprint in the expressions for
the prototype graphs 5a and 5b in ref.\cite{prototype}. The functions
$D2(3,x)$ should read $D2(2,x)$ for $x=1 \cdots 3$.
The FORTRAN files in the package `chipackage' available
on the Web are correct.}

\begin{eqnarray}
\Delta {\cal Q}_1^{L\{a\}} &=&
 \frac{g^2}{16\pi^2} \frac{s}{2\sqrt{2}} \lambda^i_{\tilde{e}e\chi}
\lambda^a_{\tilde{e}\nu\chi}
 \Bigg\{ m_{\chi_a} O^R_{ia} \Big[ -D1^{\{a\}}(2)+D2^{\{a\}}(2,1)
\nonumber \\ & & \hspace*{3 cm}
 +D2^{\{a\}}(2,3)-2D2^{\{a\}}(2,2) \Big]
\nonumber \\ & &
 +  m_{\chi_i} O^L_{ia}\Big[D0^{\{a\}}(2)-D1^{\{a\}}(1)+2D1^{\{a\}}(2)
+D1^{\{a\}}(3)
\Big] \Bigg\}, \nonumber \\
\Delta {\cal Q}_2^{L\{a\}} &=& -
 \frac{g^2}{16\pi^2} \frac{s}{2\sqrt{2}} \lambda^i_{\tilde{e}e\chi}
\lambda^a_{\tilde{e}\nu\chi}
 \Bigg\{ m_{\chi_a} O^R_{ia} \Big[ D1^{\{a\}}(2)+D2^{\{a\}}(2,1)
\nonumber \\ & & \hspace*{3 cm}
 +D2^{\{a\}}(2,3)
\Big] \Bigg\}, 
\nonumber \\
\Delta {\cal Q}_{51}^{L\{a\}} &=& 
 \frac{g^2}{16\pi^2} \frac{s}{4\sqrt{2}} \lambda^i_{\tilde{e}e\chi}
\lambda^a_{\tilde{e}\nu\chi}
 \Bigg\{ m_{\chi_a} O^L_{ia} \Big[ D0^{\{a\}}+D1^{\{a\}}(1)
 -D1^{\{a\}}(3)
\Big] \Bigg\}, \end{eqnarray}
where $g O^{R(L)}_{ia}$ are the right- (left-) handed couplings of the
$W$ to a chargino of species $a$ and a neutralino of species $i$
(see \cite{haberandkane}), $\lambda^i_{\tilde{e}e\chi}$
is the coupling between a left-handed selectron an electron and a neutralino
of species $i$, and
 $\lambda^a_{\tilde{e}\nu\chi}$
is the coupling between a left-handed selectron a neutrino and a chargino
of species $a$.   $ m_{\chi_i}$ is the neutralino mass and
 $m_{\chi_a}$ is the chargino mass. The superscript $\{a\}$ on the
Veltman-Passarino functions, $D^{\{a\}}$, indicates that they take
arguments
$$ \left( t,u, m_{\chi_a}^2, 0,m_{\chi_a}^2,0, m_{\chi_i}^2
 M_W^2,0,m_{\tilde{e}_L}^2 \right) $$

Similarly from Fig.\ref{CPfig2}(b), we obtain
\begin{eqnarray}
\Delta {\cal Q}_1^{L\{b\}} &=& 
 \frac{g^2}{16\pi^2} \frac{s}{2\sqrt{2}} \lambda^i_{\tilde{e}e\chi}
\lambda^a_{\tilde{e}\nu\chi}
 \Bigg\{ m_{\chi_a} O^R_{ia} \Big[ D1^{\{b\}}(1)-2D1^{\{b\}}(2)+D1^{\{b\}}(3)
\nonumber \\ & & \hspace*{3 cm} 
+ D2^{\{b\}}(2,1) -2D2^{\{b\}}(2,2) +D2^{\{b\}}(2,3)
\Big] \Bigg\}, 
\nonumber \\
\Delta {\cal Q}_2^{L\{b\}} &=& -
 \frac{g^2}{16\pi^2} \frac{s}{2\sqrt{2}} \lambda^i_{\tilde{e}e\chi}
\lambda^a_{\tilde{e}\nu\chi}
 \Bigg\{ m_{\chi_a} O^R_{ia} \Big[ D1^{\{b\}}(1)+D1^{\{b\}}(3)
\nonumber \\ & & \hspace*{4 cm}
 +D2^{\{b\}}(2,1)+D2^{\{b\}}(2,3) \Big]
\nonumber \\ & & \hspace*{4 cm}
 +  m_{\chi_i} O^L_{ia}\Big[D1^{\{b\}}(1)
+D1^{\{b\}}(3)
\Big] \Bigg\}, \nonumber \\
\Delta {\cal Q}_{51}^{L\{b\}} &=& 
 \frac{g^2}{16\pi^2} \frac{s}{4\sqrt{2}} \lambda^i_{\tilde{e}e\chi}
\lambda^a_{\tilde{e}\nu\chi}
 \Bigg\{ m_{\chi_a} O^L_{ia} \Big[ D0^{\{b\}}+D1^{\{b\}}(1)
 -D1^{\{b\}}(3)
\Big] \Bigg\}. \end{eqnarray}

In this case the superscript $\{b\}$
on the Veltman-Passarino functions indicates that the arguments
are
$$ \left( t,u, m_{\chi_a}^2, 0,m_{\chi_a}^2,0,
     m_{\tilde{e}_L}^2,m_{\chi_i}^2,M_W^2 \right) $$
Relations between the Veltman-Passarino functions with superscripts
  $\{a\}$ and  $\{b\}$ can be obtained by rerouting the loop
momentum in one of the graphs. When this is effected, we find the relations
$$ \Delta {\cal Q}_2^{L\{b\}} \ = \ \Delta {\cal Q}_1^{L\{a\}} $$
$$ \Delta {\cal Q}_2^{L\{a\}} \ = \ \Delta {\cal Q}_1^{L\{b\}} $$
$$ \Delta {\cal Q}_{51}^{L\{b\}} \ = \ -\Delta {\cal Q}_{51}^{L\{a\}} $$
so that the sum of the contributions from the two graphs satisfies the 
relations (\ref{CP1}), as required. We have not implemented these 
relations, but rather checked numerically that the CP invariance 
relations are obeyed, thus providing a stringent check on the numerical 
computation of the Veltman-Passarino functions. 

We note that CP invariance {\it cannot} be used to relate the differential
cross-sections
$\sigma^{+-}$ and $\sigma^{-+}$. Differences between these
reflect the presence of parity violating interactions.

\setlength{\unitlength}{0.29mm}

\section{Results}

To present our results we have chosen four benchmark points, defined in
ref.~\cite{benchmark} as benchmark points C, E, I, and L. One of them,
benchmark point C, has been included in the Snowmass 2001 benchmark models
as SPS3 \cite{snowmassBP}. Details about the four benchmark points are 
given in the references above. Here, it is in our interest to mention the
values of the parameters which 
are relevant for
 the calculation of the 
cross-sections at tree level, and to give some typical values of the masses 
of the particles which 
are relevant at one loop. For benchmark C we have
at one--loop
\begin{equation}
M=315 \,{\mathrm{GeV}}, \quad \mu=494 \,{\mathrm{GeV}}, \quad
\tan\beta=10,  \quad m_{\tilde\nu_e}=279 \,{\mathrm{GeV}}.
\label{oneloopparam}
\end{equation}
which generates the following chargino masses
\begin{equation}
m_{\tilde\chi_1^+}=310 \,{\mathrm{GeV}}, \qquad
m_{\tilde\chi_2^+}=533 \,{\mathrm{GeV}}.
\end{equation}
Radiative corrections change the value of the chargino masses compared with 
their tree-level value\cite{PP,mass1loop}. Nevertheless, in this paper we 
will compare the radiatively corrected chargino production cross-sections 
with tree-level cross-sections calculated with values of the parameters
$M$ and $\mu$ modified such that the chargino masses are equal in both 
cases. In other words, we are fixing the chargino masses and 
transplanting the
effect of quantum corrections to $M$ and $\mu$. For benchmark C the 
tree-level parameters are
\begin{equation}
M^{(0)}=326 \,{\mathrm{GeV}}, \quad \mu^{(0)}=511 \,{\mathrm{GeV}}, 
\label{treelevelparam}
\end{equation}
which amounts to a $3.4\%$ correction for each parameter, between
the values taken for the tree-level calculation and those taken for the
one-loop calculation. Unless otherwise stated, we use a center of mass 
energy of $\sqrt{s}=1$ TeV.

\begin{figure}
\centerline{\protect\hbox{\epsfig{file=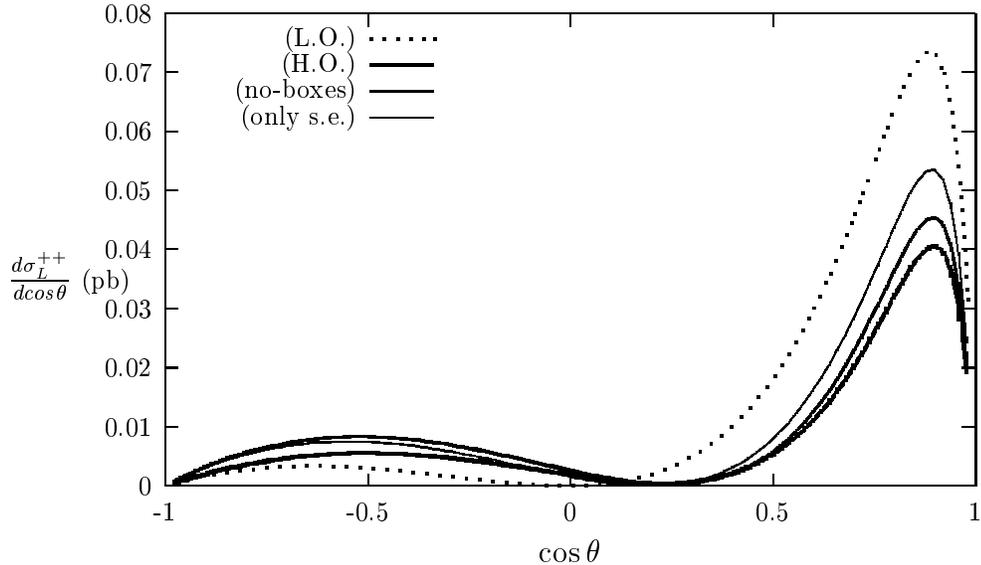,height=7.5cm,width=0.85
 \textwidth}}}
\caption{\it Lowest order, complete higher order, no--boxes, and 
only--self--energy cross-sections for positive helicity light chargino 
pair production for left-polarized electrons in benchmark C model.} 
\label{Lpp_Ccom} 
\end{figure} 
In order to compare the relative importance of self-energy, triangle, and 
box diagrams we have plotted in Fig.~\ref{Lpp_Ccom} the differential cross
section $\sigma_L^{++}$ for the production of positive helicity light 
charginos with left handed electrons (and right handed positrons) as a 
function of the scattering angle.
We show the tree-level cross-section, calculated with parameters in 
eq.~(\ref{treelevelparam}), the complete one--loop cross-section calculated
with the parameters indicated in eq.~(\ref{oneloopparam}), the one-loop
cross-section where the contribution from box diagrams has been removed, 
and the one--loop cross-section where only self-energy diagrams have been
included. In this case, at small angles where the cross-section is larger,
self-energy contributions are somewhat more than half of the total, while
the contribution form triangles is about half of that, and similarly for 
boxes \footnote{
We note that the super-oblique corrections are generated by self-energy
type of diagrams \cite{superoblique}. 
}. This confirms previous statements about the importance of triangles 
\cite{DKR,KNPY,DKR2} and boxes \cite{BH}. At large angles, where the cross 
section is smaller, the importance of boxes is even larger.

\begin{figure}
\centerline{\protect\hbox{\epsfig{file=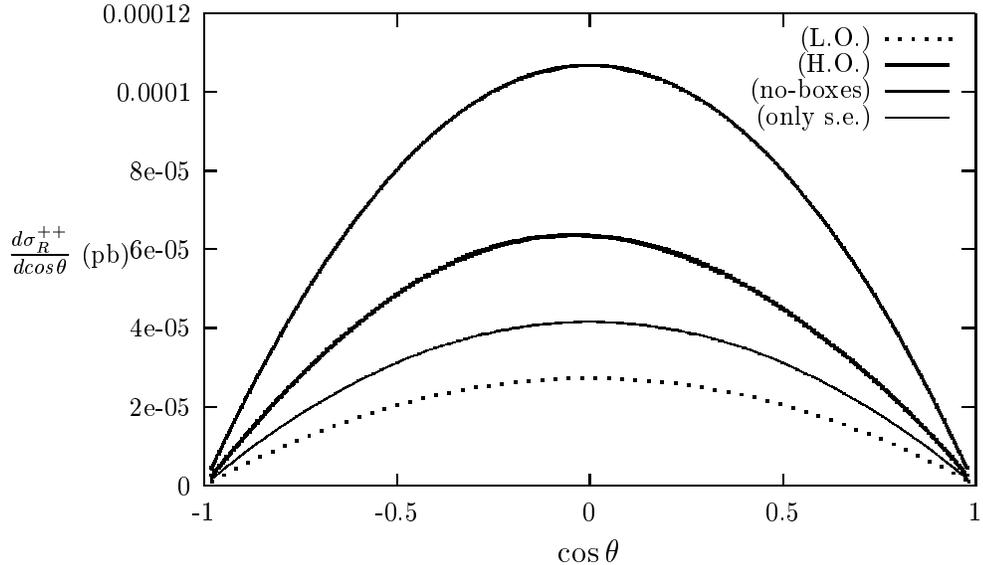,height=7.5cm,width=0.85
\textwidth}}}
\caption{\it Lowest order, complete higher order, no--boxes, and 
only--self--energy cross-sections for positive helicity light chargino 
pair production for right-polarized electrons in benchmark C model.
} 
\label{Rpp_Ccom} 
\end{figure} 
Plotted in Fig.~\ref{Rpp_Ccom} is the production cross-section 
$\sigma_R^{++}$ for positive helicity light charginos with right handed 
electrons (and left handed positrons). It is much smaller than the left 
handed case, but probably still observable in a future linear collider. 
In this case, the
self--energy contribution is the smallest and the inclusion of triangles 
and boxes is crucial. Note that despite the partial cancellation between
triangle and box contributions, the net effect of quantum corrections is to 
double the cross-section.

\begin{table}
\centering
\renewcommand{\arraystretch}{0.95}
{~}\\
\begin{tabular}{|c||r|r|r|r|}
\hline
Model             & C   &  E  &  I  &  L   \\ \hline
Parameters        &     &     &     &      \\ \hline
$\tan{\beta}$     & 10  & 10  & 35  & 50   \\
$M$ (GeV)         & 315 & 249 & 276 & 359  \\
$\mu$ (GeV)       & 494 & 230 & 437 & 544  \\ \hline
$M^{(0)}$ (GeV)   & 326 & 249 & 285 & 372  \\
$\mu^{(0)}$ (GeV) & 511 & 253 & 456 & 580  \\ \hline
Masses (GeV)      &     &     &     &      \\ \hline
$\chi^{\pm}_1$    & 310 & 194 & 271 & 360  \\
$\chi^{\pm}_2$    & 533 & 318 & 478 & 598  \\ \hline
$\tilde\nu_e$     & 279 &1512 & 292 & 459  \\ \hline
A$^0$             & 576 &1509 & 449 & 491  \\
$t_1$             & 612 &1029 & 573 & 784  \\
$b_1$             & 759 &1354 & 646 & 811  \\
\hline
\end{tabular}
\caption[]{\it Four models, motivated by the benchmark points C, E, I, 
and L described in \cite{benchmark,snowmassBP}, used for the calculation 
of the production cross-section of two light charginos with definite 
helicity in $e^+e^-$ annihilation with polarized beams.
As explained in section \ref{intro}, these parameters are understood
 to be in the $\overline{DR}$ scheme. } 
\label{tab:benchmark}
\end{table}
In what follows we compare the complete one--loop corrected chargino pair 
production cross-section with the tree-level cross-section calculated in 
the way described above, working with four benchmark points. In table 
\ref{tab:benchmark} we describe the four models including the most 
relevant parameter and masses. In all models we have a reasonably light
Higgs mass satisfying $m_h>112$ GeV, and values of $b\rightarrow s \gamma$
within the 95\% confidence limit 
$2.33\times 10^{-4}<{\cal B}(b\rightarrow s \gamma)<4.15\times 10^{-4}$.
Values of the $\tilde t_1$, $\tilde b_1$, and $m_A$ are given in order to 
have an idea of the squark and heavy Higgs boson masses, which intervene 
logarithmically in the corrections.

\begin{figure}
\centerline{\protect\hbox{\epsfig{file=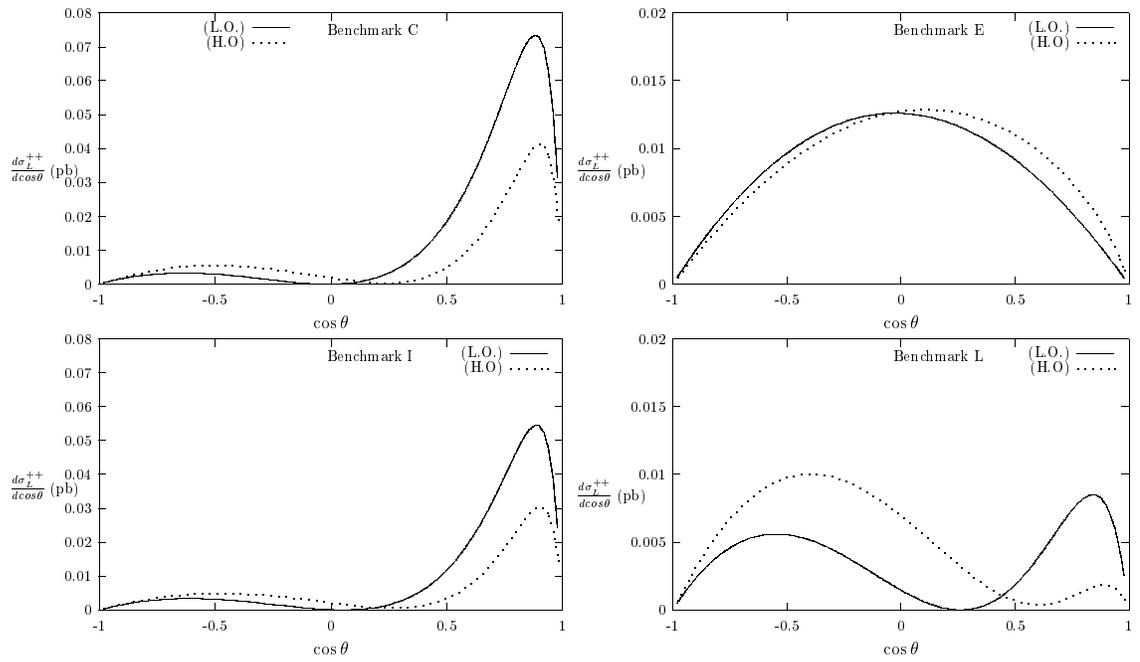,width=15cm}}}
\caption{\it Lowest and higher order cross-section $\sigma_L^{++}$ for the
production of two light charginos with left-polarized electrons and positive
helicity for both charginos.
} 
\label{siglpp} 
\end{figure} 
In Fig.~\ref{siglpp} we plot the differential cross-section $\sigma_L^{++}$
for the production of a pair of positive helicity light charginos with a 
beam of left handed electrons. The four benchmark points are displayed in 
each frame. The differential cross-sections for benchmarks C and I have a 
similar shape, with a large forward-backward asymmetry. In both cases the 
cross-section decreases after adding quantum corrections. The differential 
cross-section for benchmark E is much more symmetric and radiative 
corrections are small. Note that in this case the sneutrino mass is 1.5 TeV 
and its t--channel contribution is suppressed. In benchmark point L there is 
also a large correction to the forward-backward asymmetry. In addition, the 
angle at which the tree-level cross-section is zero is corrected quite 
substantially at one loop.

\begin{figure}
\centerline{\protect\hbox{\epsfig{file=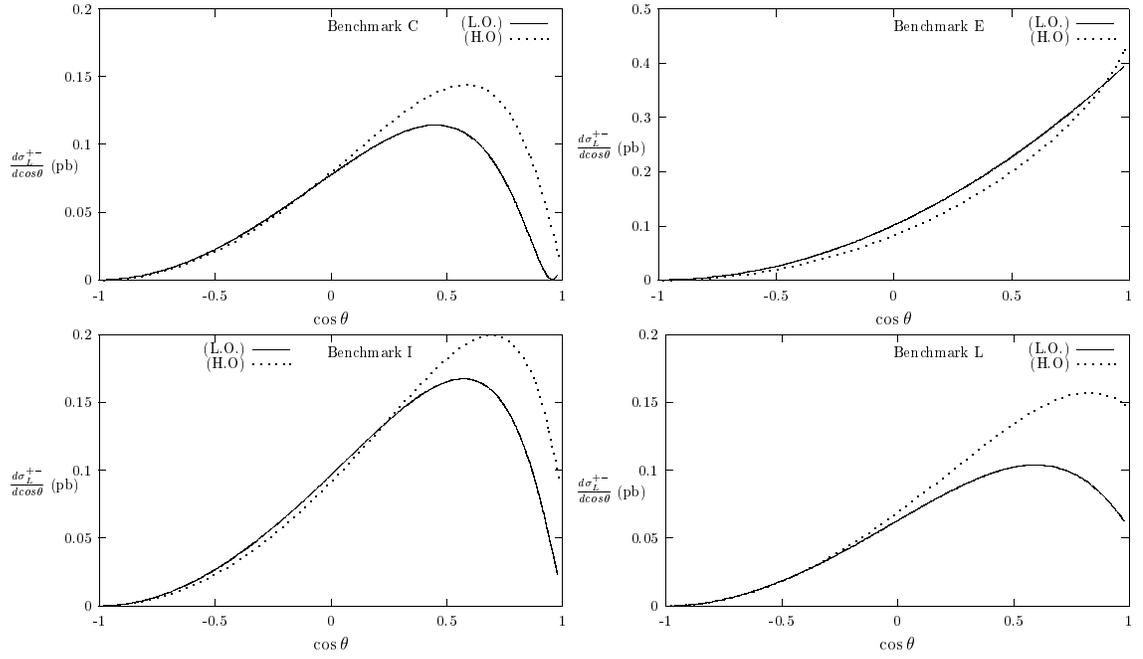,width=15cm}}}
\caption{\it Lowest and higher order cross-section $\sigma_L^{+-}$ for the
production of two light charginos with left-polarized electrons and positive
(negative) helicity for the chargino (anti-chargino).
}
\label{siglpm} 
\end{figure} 
In Fig.~\ref{siglpm} we have $\sigma_L^{+-}$, corresponding to the production
of a positive helicity chargino and a negative helicity anti-chargino with
left handed electrons. In this case, benchmark C, I, and L are similar and
have a large forward-backward asymmetry, and in the three cases the cross 
sections and $A_{FB}$ increase after the inclusion of radiative corrections. 
As before, the radiative corrections for benchmark E are small.

\begin{figure}
\centerline{\protect\hbox{\epsfig{file=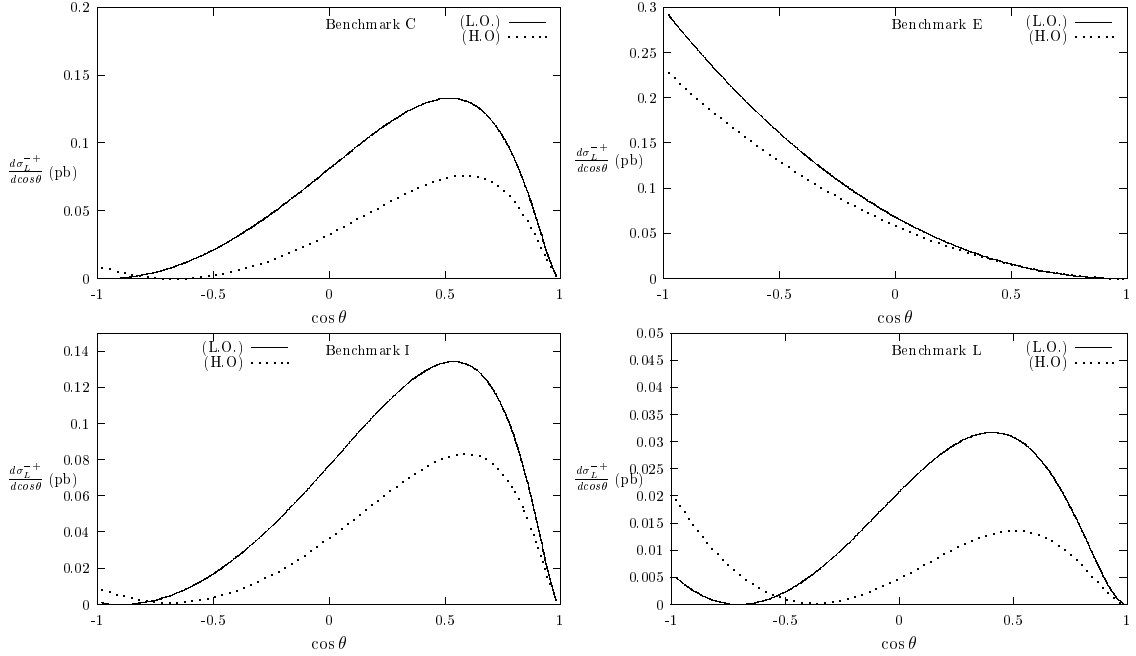,width=15cm}}}
\caption{\it Lowest and higher order cross-section $\sigma_L^{-+}$ for the
production of two light charginos with left-polarized electrons and negative
(positive) helicity for the chargino (anti-chargino).
} 
\label{siglmp}
\end{figure}

In Fig.~\ref{siglmp} we plot $\sigma_L^{-+}$, where a negative helicity 
chargino and a positive helicity anti-chargino are produced with left handed
electrons. Contrary to the previous case, corrections to the cross-section
for benchmarks C, I, and L are negative, and particularly in the later case, 
affecting importantly the forward-backward asymmetry. Corrections in 
benchmark E are negative and large at backward angles.
\begin{figure}
\centerline{\protect\hbox{\epsfig{file=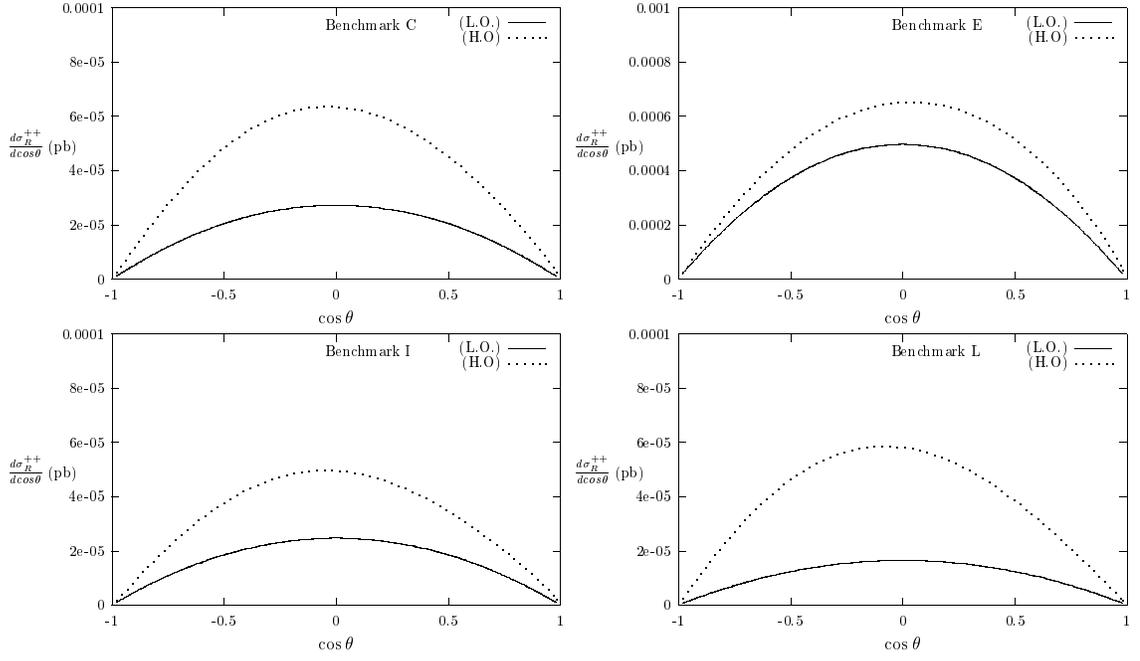,width=15cm}}}
\caption{\it Lowest and higher order cross-section $\sigma_R^{++}$ for 
the production of two light charginos with right-polarized electrons and 
positive helicity for both charginos.}
\label{sigrpp}
\end{figure} 
In the following three figures we consider cross-sections for right handed
electrons. In Fig.~\ref{sigrpp} we plot $\sigma_R^{++}$ corresponding to the 
production of two charginos with positive helicity. In all cases the 
differential cross-section is maximal at $\cos\theta=0$ with a very small
forward-backward asymmetry. Radiative corrections are very large in all cases,
specially for benchmark L where the cross-section increases
 several times.

\begin{figure}
\centerline{\protect\hbox{\epsfig{file=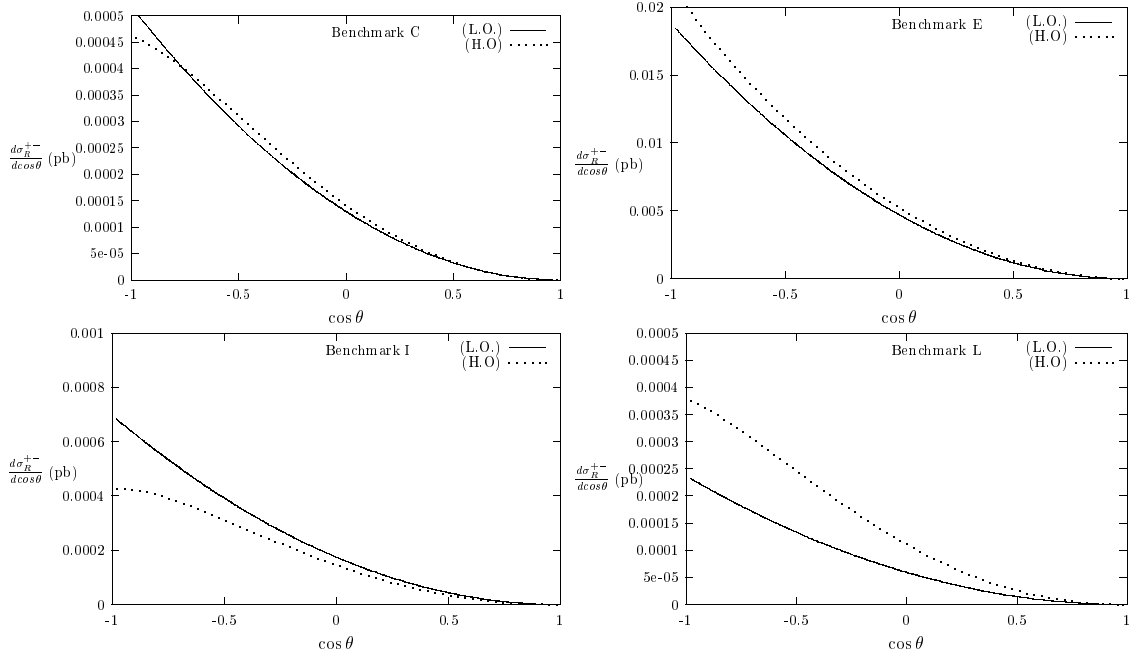,width=15cm}}}
\caption{\it Lowest and higher order cross-section $\sigma_R^{+-}$ for 
the production of two light charginos with right-polarized electrons and 
positive (negative) helicity for the chargino (anti-chargino).
}
\label{sigrpm}
\end{figure}

The differential cross-section $\sigma_R^{+-}$ is plotted in Fig.~\ref{sigrpm}
as a function of the scattering angle, corresponding to the production of 
a negative helicity chargino and a positive helicity anti-chargino with 
right handed electrons. For all benchmark points the cross-section is 
maximum at large angles, and corrections are large for benchmark L, and 
non-negligible for the rest.
\begin{figure}
\centerline{\protect\hbox{\epsfig{file=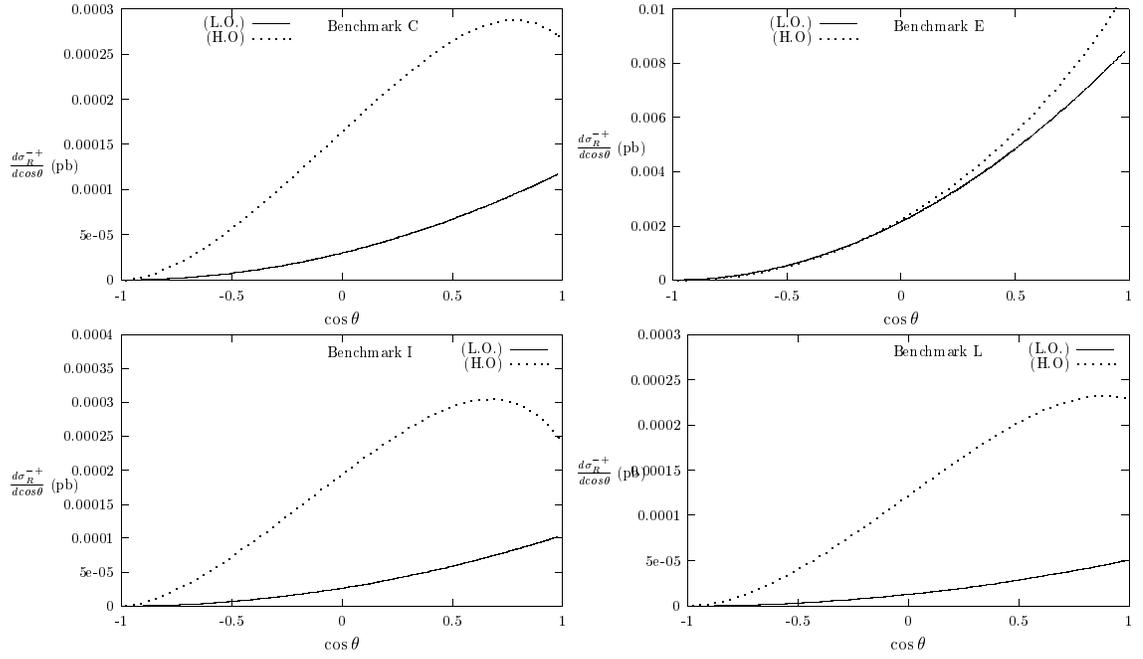,width=15cm}}}
\caption{\it Lowest and higher order cross-section $\sigma_R^{-+}$ for 
the production of two light charginos with right-polarized electrons and 
negative (positive) helicity for the chargino (anti-chargino).}
\label{sigrmp}
\end{figure} 
Finally, in Fig.~\ref{sigrmp} we have $\sigma_R^{-+}$ corresponding to the 
differential cross-section for the production of a negative helicity 
chargino and a positive helicity anti-chargino. In all cases the cross 
section is maximal at small angles. Huge corrections are found for 
benchmark points C, I, and L, where the total cross-section increase
several times due to the higher order corrections.

To finish this section, we compare with the results published in 
ref.~\cite{BH}, where a complete one-loop calculation of the chargino pair
production cross section was performed, although without projecting
the helicities of the final-state charginos.

\begin{figure}
\centerline{\protect\hbox{\epsfig{file=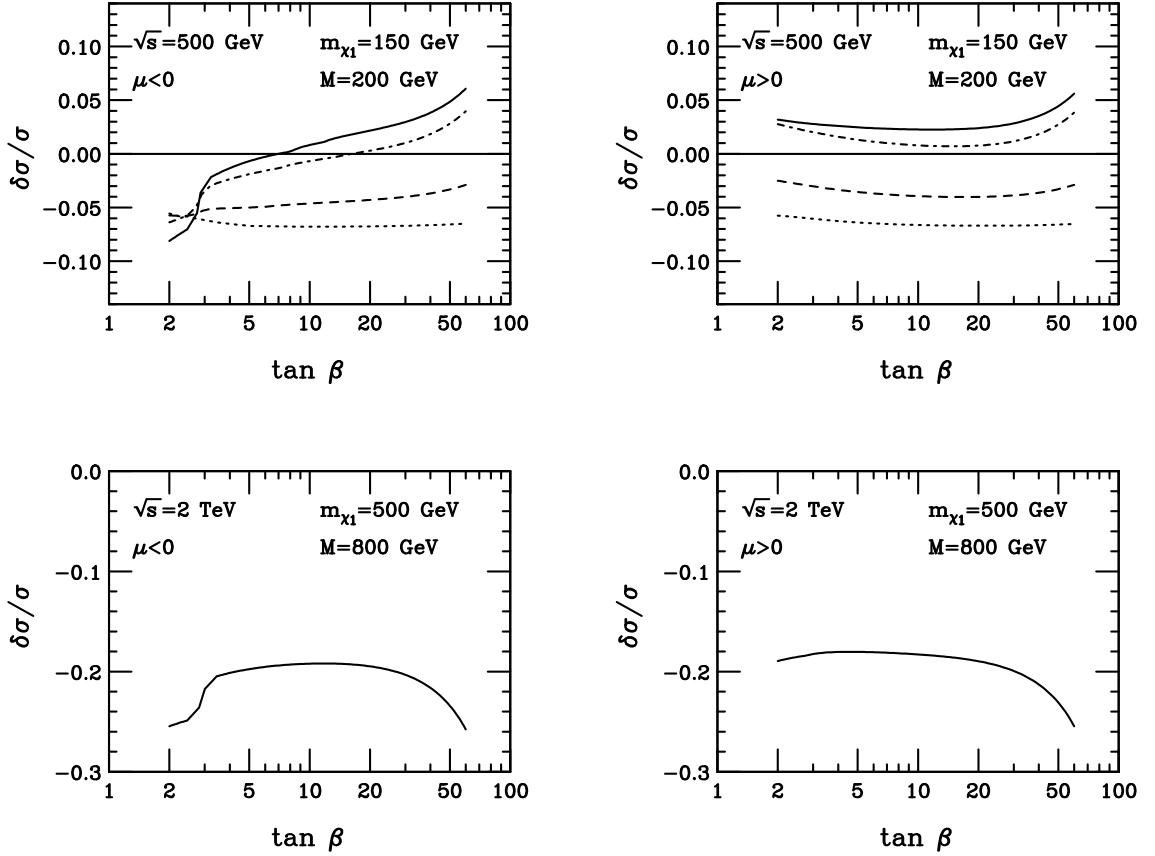,width=0.75
\textwidth,angle=90}}}
\caption{
Quantum corrections to the unpolarized chargino pair production cross
section as a function of $\tan\beta$ for different cases as explained in 
the text.
} 
\label{BH_CS} 
\end{figure} 
In Fig.~\ref{BH_CS} we plot the relative correction to the unpolarized 
total cross section $\delta\sigma/\sigma\equiv(\sigma_1-\sigma_0)/\sigma_0$.
Here $\sigma_1$ is the one loop unpolarized cross section calculated with a 
fixed value of the gaugino mass $M$ (200 GeV for the upper quadrants and 800
GeV for the lower quadrants) and a value of $\mu^{(1)}$ such that the light 
chargino mass is fixed (150 GeV and 500 GeV respectively). Similarly, 
$\sigma_0$ is the tree level cross section calculated with the same value 
of the gaugino mass $M$ and a different value of the higgsino mass 
$\mu^{(0)}$ such that we obtain the same numerical value for the light 
chargino mass at tree level. 

\begin{figure}
\centerline{\protect\hbox{\epsfig{file=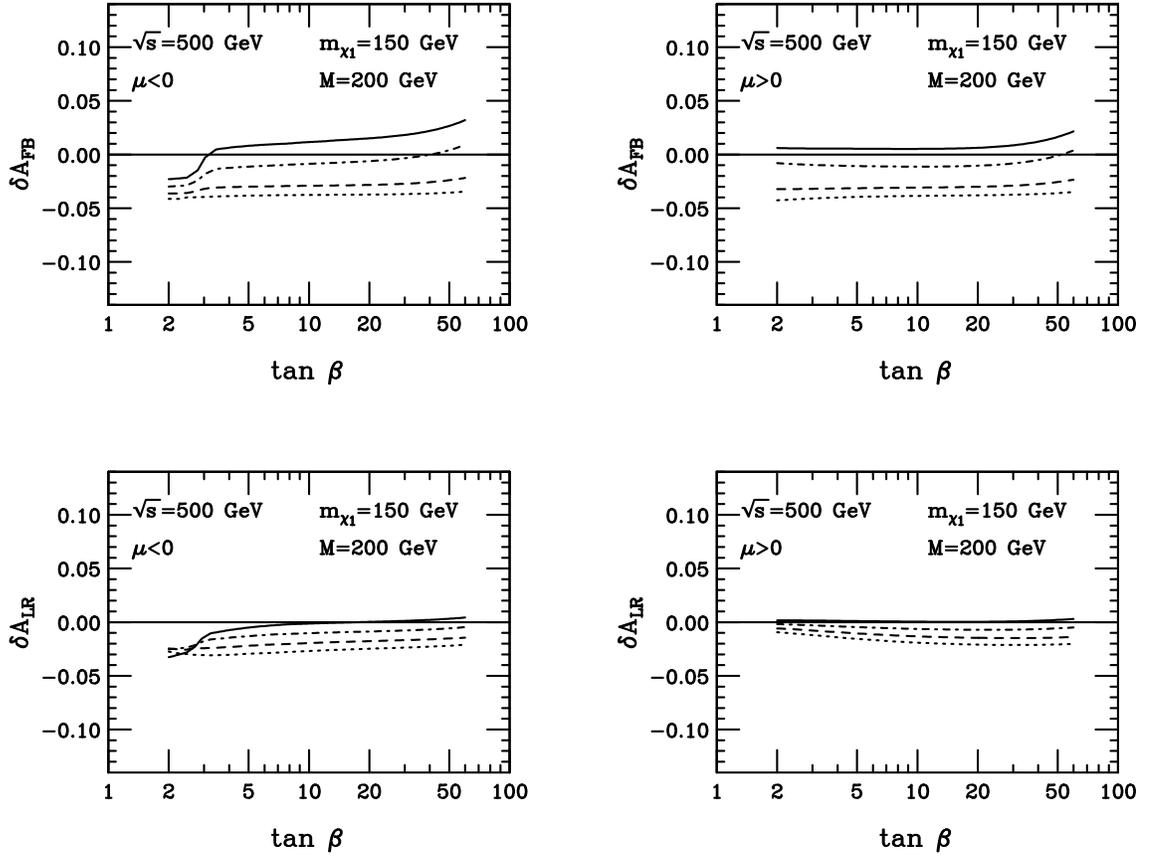,width=0.75
\textwidth,angle=90}}}
\caption{
Quantum corrections to the forward-backward and left-right asymmetries
as a function of $\tan\beta$ for different cases as explained in 
the text.
} 
\label{BH_A} 
\end{figure} 
In the upper quadrants of Fig.~\ref{BH_CS} we take a center of mass energy of
$\sqrt{s}=500$ GeV, a light chargino mass of $m_{\chi_1^+}=150$ GeV, and
a gaugino mass of $M=200$ GeV. The quadrants at the left (right) correspond
to $\mu<0$ ($\mu>0$). We take all the squark and slepton soft mass parameters
degenerate at the weak scale and equal to $M_{SUSY}$. Four curves are shown
corresponding to $M_{SUSY}=300$ GeV (solid), 500 GeV (dotdash), 1 TeV
(dash), and 2 TeV (dots). We plot them as a function of $\tan\beta$ in an 
interval motivated by a correct radiative electroweak symmetry breaking
and the LEP exclusion of values of $\tan\beta$ close to unity.
We see that corrections are of the order of $5\%$ and become more
negative as $M_{SUSY}$ is increased. In addition, for $\mu<0$ the four 
curves tend to focus near $-5\%$ at low $\tan\beta$ and spread out at large
$\tan\beta$, as opposed to the case with $\mu>0$ where there is a milder
dependence on $\tan\beta$. These general features are observed also in 
Fig.1 of ref.~\cite{BH}. There is, however, 
some quantitative difference between 
our results and those of ref.~\cite{BH}, particularly at low values of
$M_{SUSY}$, where our corrections are more positive than theirs.

In the lower quadrants of Fig.\ref{BH_CS} we take a center of mass energy of
$\sqrt{s}=2$ TeV, a light chargino mass of $m_{\chi_1^+}=500$ GeV, and
a gaugino mass of $M=800$ GeV. The corrections are larger than the previous
case and of the order of $-20\%$ and becoming more negative for extreme 
values of $\tan\beta$. The same behaviour is observed in
Fig 2. of  ref.\cite{BH}, where they also report large negative corrections
specially at extreme values of $\tan\beta$.

In Fig.~\ref{BH_A} we concentrate on the forward-backward asymmetry $A_{FB}$
and the left-right asymmetry $A_{LR}$. In the upper quadrants we plot the
extra contribution of quantum corrections to the forward-backward asymmetry
$\delta A_{FB}\equiv A_{FB}^1-A_{FB}^0$. In the lower quadrants we plot the
extra contribution of quantum corrections to the left-right asymmetry
$\delta A_{LR}\equiv A_{LR}^1-A_{LR}^0$. In all cases we consider a center 
of mass energy of $\sqrt{s}=500$ GeV, a light chargino mass of 
$m_{\chi_1^+}=150$ GeV, and a gaugino mass of $M=200$ GeV. Corrections
to $A_{FB}$ are less that $5\%$ and negative for large values of $M_{SUSY}$,
and a stronger dependence on $\tan\beta$ for $\mu<0$. This behaviour is 
also observed in Fig.4 of ref.~\cite{BH}. The  quantitative differences 
are as before, 
i.e. a stronger dependence on $M_{SUSY}$ and corrections displaced towards
 the positive side  for low values of $M_{SUSY}$ in our case.

In the lower quadrants of Fig.~\ref{BH_A} we plot the corrections to
$A_{LR}$. They are smaller than $3\%$ for $\mu<0$ and smaller than $2\%$
for $\mu>0$. In both cases they are negative. For $\mu<0$ a small positive 
slope is observed for the curves as $\tan\beta$ increases, and for $\mu>0$
a small negative slope is present. As before, this general behaviour is
also seen in Fig.5 of ref.~\cite{BH}. The difference is again a larger 
dependence on $M_{SUSY}$ in our case.

As seen from the last two figures, we find moderate corrections to the 
production cross section ($5\%$) and to the $A_{FB}$ ($4\%$) and $A_{LR}$ 
($3\%$) asymmetries for the scenario with $\sqrt{s}=500$ GeV, and 
larger corrections to the cross section ($20\%$) for the scenario with
$\sqrt{s}=2$ TeV. These magnitudes are in agreement with the corrections 
found in the same scenarios in ref.~\cite{BH}. Several other features and 
dependence on parameters of the corrections are present in our plots as 
well as in the plots in ref.~\cite{BH}.

The modest quantitative disagreement between the results shown here
and those of ref.~\cite{BH} is almost certainly due to the difference
of renormalization prescription.  In ref.~\cite{BH}  `on-shell'
 renormalization is used, but unfortunately the exact
processes that 
are used to determine the various renormalized parameters, are not specified. 
It is therefore not possible to make any further comparison. It is perfectly
plausible that sensitivity to the values of the SUSY parameters
at the few percent level can arise from a corresponding sensitivity of
the translation between on-shell renormalized parameters and those
in the $\overline{DR}$. On the other hand, one should note that the
very dramatic effects of higher order corrections arise (particularly in
the cross-sections for right- polarized electrons) because a given process
 - or contribution to a process - is forbidden at tree-level, but permitted
at one-loop level through box-graphs. A change in renormalization 
prescription can only affect contributions to processes that are permitted
at tree-level. We do not therefore anticipate that the very large corrections
reported here would be significantly altered by a change in the renormalization
prescription.

\section{Conclusions}

We have calculated the complete weak one--loop corrections to the production
of two charginos in electron--positron colliders. We consider the 
polarization of the electron and positron, and the helicity of the 
charginos. We include all self-energy, triangle, and box diagrams of
weak interaction in the MSSM, leaving out the calculation of the QED
contributions which will be addressed elsewhere. Confirming previous
calculations we find that triangle and boxes cannot be neglected, and in
some cases are even dominant. We have displayed the radiative corrections 
to the different differential cross-sections in four benchmark points,
chosen in ~\cite{benchmark} as representative scenarios for supersymmetry,
one of them included in the Snowmass 2001 benchmark models \cite{snowmassBP}.
The correction we found are usually very large (tens of percent) and 
sometimes they are huge (hundreds of percent). If charginos are discovered,
for example at the LHC, the underlying parameters of the model can only be
extracted through precision measurements at a future $e^+e^-$ Linear Collider,
and our results indicate that this program can only be carried successfully
if full one-loop corrections are included.

\newpage
\noindent{\bf Acknowledgements:}

\noindent
This work was partially supported by CONICYT grant No.~1010974.

\vskip1cm

\section*{Appendix}

In this appendix we show all the one-loop diagrams included in this 
calculation, separating them in self energies, triangles, and boxes.
For each prototype diagram defined in \cite{prototype}, we indicate the
internal particles which define each loop.

\subsection*{Self-energies}

\noindent {\bf Sneutrino Self-Energy}

\begin{center}
\begin{picture}(333,125)
\SetScale{0.8}
\footnotesize{
\DashLine(0,75)(100,75){6}
\DashLine(0,75)(100,75){6}
\DashLine(200,75)(300,75){6}
\PhotonArc(150,75)(50,0,180){4}{10}
\DashCArc(150,75)(50,180,360){6}  
 } \put(99,5){Self-energy prototype 2}
\end{picture}
\end{center}
The possible internal particles are:
$$ Z,\  \tilde{\nu} $$ 
$$ W, \ \tilde{e}_L $$  
\bigskip

\begin{center}
\begin{picture}(333,125)
\SetScale{0.8}
\footnotesize{
\DashLine(0,75)(100,75){6}
\DashLine(200,75)(300,75){6}
\DashCArc(150,75)(50,0,180){6}
\DashCArc(150,75)(50,180,360){6}
}  \put(99,5){Self-energy prototype 3}
\end{picture}
\end{center}
The possible internal particles are:
$$ H^0 (\phi^0), \  \tilde{\nu} $$ 
$$ H^\pm (\phi^\pm), \  \tilde{e}_L $$ 
Here and below, $H^\pm(\phi^\pm)$ stands for all possible charged 
Higgs particles or the $W$-Goldstone bosons  and similarly $H^0(\phi^0)$
stands for the neutral Higgs scalar, the pseudo-scalar and the $Z$-Goldstone
boson. 
\bigskip

\begin{center}
\begin{picture}(333,125)
\SetScale{0.8}
\footnotesize{
\DashLine(0,75)(100,75){6}
\DashLine(200,75)(300,75){6}
\ArrowArc(150,75)(50,0,180)
\ArrowArc(150,75)(50,180,360)
}  \put(99,5){Self-energy prototype 4}
\end{picture}
\end{center}
The possible internal particles are:
$$ \tilde{\chi}^0_a, \   \nu, \ \  (a=1 \cdots 4)  $$ 
$$ \tilde{\chi}^+_a, \   e, \ \  (a=1 \cdots 2)  $$  
\bigskip

\noindent {\bf Gauge-Boson Self-Energy}

\begin{center}
\begin{picture}(333,125)
\SetScale{0.8}
\footnotesize{
\Photon(0,75)(100,75){4}{10}
\Photon(200,75)(300,75){4}{10}
\PhotonArc(150,75)(50,0,180){4}{10}
\PhotonArc(150,75)(50,180,360){4}{10}
} \put(99,5){Self-energy prototype 5}
\end{picture}
\end{center}
The internal particles are $W^+, \ W^-$. 
\bigskip

\begin{center}
\begin{picture}(333,125)
\SetScale{0.8}
\footnotesize{
\Photon(0,75)(100,75){4}{10}
\Photon(200,75)(300,75){4}{10}
\PhotonArc(150,75)(50,180,360){4}{10}
\DashCArc(150,75)(50,0,180){6}
} \put(99,5){Self-energy prototype 6}
\end{picture}
\end{center}
The possible internal particles are:
$$ H^0(\phi^0), \ Z $$
$$ H^\pm(\phi^\pm), \ W $$ 
This diagram
includes the contribution from Faddeev-Popov ghosts
as well as the accompanying tadpole graph involving the
quartic gauge-boson coupling.
\bigskip

\begin{center}
\begin{picture}(333,125)
\SetScale{0.8}
\footnotesize{
\Photon(0,75)(100,75){4}{10}
\Photon(200,75)(300,75){4}{10}
\DashCArc(150,75)(50,0,180){6}
\DashCArc(150,75)(50,180,360){6}
} \put(99,5){Self-energy prototype 7}
\end{picture}
\end{center}
The possible internal particles are:
$$ H^0(\phi^0), \ H^0(\phi^0 $$
$$ H^\pm(\phi^\pm), \   H^\mp(\phi^\mp)$$ 
$$ \tilde{f}, \ \tilde{f} $$
where $\tilde{f}$ stands for all the scalar super-partners associated
with matter fermions (lepton and quarks of both helicities).
\bigskip

\begin{center}
\begin{picture}(333,125)
\SetScale{0.8}
\footnotesize{
\Photon(0,75)(100,75){4}{10}
\Photon(200,75)(300,75){4}{10}
\ArrowArc(150,75)(50,0,180)
\ArrowArc(150,75)(50,180,360)
}  \put(99,5){Self-energy prototype 8}
\end{picture}
\end{center}
The possible internal particles are:
$$ \tilde{\chi}^0_a, \ \tilde{\chi}^0_b, \ \ (a,b=1 \cdots 4) $$
$$ \tilde{\chi}^-_a, \ \tilde{\chi}^+_b, \ \ (a,b=1 \cdots 2) $$
$$ f, \ f$$
where $f$ stands for all the matter fermions.
\bigskip

\noindent {\bf Chargino Self-Energy}

\begin{center}
\begin{picture}(333,125)
\SetScale{0.8}
\footnotesize{
\ArrowLine(0,75)(100,75)
\ArrowLine(200,75)(300,75)
\ArrowArc(150,75)(50,180,360)
\PhotonArc(150,75)(50,0,180){4}{10}
}  \put(99,5){Self-energy prototype 9}

\end{picture}
\end{center}
The possible internal particles are:
$$ \tilde{\chi}^0_a, \ Z \ \ (a=1 \cdots 4) $$
$$ \tilde{\chi}^-_a, \ W, \ \ (a=1 \cdots 2) $$
\bigskip

\begin{center}
\begin{picture}(333,125)
\SetScale{0.8}
\footnotesize{
\ArrowLine(0,75)(100,75)
\ArrowLine(200,75)(300,75)
\ArrowArc(150,75)(50,180,360)
\DashCArc(150,75)(50,0,180){6}
}  \put(99,5){Self-energy prototype 10}
\end{picture}
\end{center}

The possible internal particles are:
$$ \tilde{\chi}^0_a, \ H^0(\phi^0) \ \ (a=1 \cdots 4) $$
$$ \tilde{\chi}^-_a, \ H^+(\phi^+) \ \ (a=1 \cdots 2) $$
$$ f, \ \tilde{f} $$
where $f$ stands for all the matter fermions (quarks and leptons of
both chiralities) and $\tilde{f}$ are their corresponding 
scalar super-partners.
\bigskip

\subsection*{Vertex Corrections}

\noindent {\bf Sneutrino-Electron-Chargino Vertex}

\begin{center}
\begin{picture}(333,150)
\SetScale{0.8}
\footnotesize{
\DashLine(0,75)(130,75){5}
\ArrowLine(310,25)(215,25)
\ArrowLine(215,25)(130,75)
\ArrowLine(215,125)(130,75) 
\ArrowLine(310,125)(215,125) 
\Photon(215,25)(215,125){4}{10}
\put(170,85){$1$}
\put(170,51){$3$}
\put(195,70){$2$}} \put(99,5){Triangle prototype 1}
\end{picture} 
\end{center}
The possible internal particles (ordered from 1 to 3) are:
$$ \nu, \ W, \ \tilde{\chi}^0_a,  \ \ (a=1 \cdots 4) $$ 
$$ e, \ Z, \  \ \tilde{\chi}^-_a, \ \ (a=1 \cdots 2) $$  
\bigskip

\begin{center}
\begin{picture}(333,150)
\SetScale{0.8}
\footnotesize{
\DashLine(0,75)(130,75){6}
\ArrowLine(310,25)(215,25) 
\Photon(215,25)(130,75){4}{10}
\DashLine(215,125)(130,75){6}
\ArrowLine(310,125)(215,125)
\ArrowLine(215,25)(215,125)
\put(170,85){$1$}
\put(170,55){$3$}
\put(195,70){$2$}} \put(99,5){Triangle prototype 2a}
\end{picture}
\end{center}
The possible internal particles (ordered from 1 to 3) are:
$$ \tilde{\nu}, \ \tilde{\chi}^-_a , \ Z,  \ \ (a=1 \cdots 2) $$ 
$$ \tilde{e}_L, \ \tilde{\chi}^0_a , \ W,  \ \ (a=1 \cdots 4) $$ 
\bigskip

\begin{center}
\begin{picture}(333,150)
\SetScale{0.8}
\footnotesize{
\DashLine(0,75)(130,75){6}
\ArrowLine(310,25)(215,25) 
\DashLine(215,25)(130,75){6}
\Photon(215,125)(130,75){4}{10}
\ArrowLine(310,125)(215,125) 
\ArrowLine(215,25)(215,125)
\put(170,80){$1$}
\put(170,51){$3$}
\put(195,70){$2$}}  \put(99,5){Triangle prototype 2b}
\end{picture}
\end{center}
The internal particles (ordered from 1 to 3) are:
$$ Z, \ e , \tilde{\nu}$$ 
\bigskip

\begin{center}
\begin{picture}(333,150)
\SetScale{0.8}
\footnotesize{
\DashLine(0,75)(130,75){6}
\ArrowLine(310,25)(215,25)
\DashLine(215,25)(130,75){6}
\DashLine(215,125)(130,75){6}
\ArrowLine(310,125)(215,125) 
\ArrowLine(215,25)(215,125)

\put(170,80){$1$}
\put(170,50){$3$}
\put(192,70){$2$}}  \put(99,5){Triangle prototype 3}
\end{picture}
\end{center}
The possible internal particles (ordered from 1 to 3) are:
$$ \tilde{e}_L, \ \tilde{\chi}^0_a , \ H^\pm(\phi^\pm) ,  \ \ 
(a=1 \cdots 4) $$
$$ \tilde{\nu}, \ \tilde{\chi}^-_a , \ H^0(\phi^0)  \ \ (a=1 \cdots 2) $$ 
\bigskip

\begin{center}
\begin{picture}(333,150)
\SetScale{0.8}
\footnotesize{
\DashLine(0,75)(130,75){6}
\ArrowLine(310,25)(215,25) 
\ArrowLine(215,25)(130,75)
\ArrowLine(215,125)(130,75)
\ArrowLine(310,125)(215,125) 
\DashLine(215,25)(215,125){6}
\put(170,80){$1$}
\put(170,50){$3$}
\put(192,70){$2$}} \put(99,5){Triangle prototype 4}
\end{picture}
\end{center}
The possible internal particles (ordered from 1 to 3) are:
$$ \tilde{\chi}^0_a, \ \tilde{e}_L , \ \nu ,  \ \ (a=1 \cdots 4) $$
$$ \tilde{\chi}^-_a, \ \tilde{\nu} , \ e ,  \ \ (a=1 \cdots 2) $$
\bigskip

\noindent {\bf Gauge Boson-Chargino-Chargino Vertex}

\begin{center}
\begin{picture}(333,150)
\SetScale{0.8}
\footnotesize{
\Photon(0,75)(130,75){4}{12}
\ArrowLine(310,25)(215,25) 
\ArrowLine(215,25)(130,75)
\ArrowLine(130,75)(215,125)
\ArrowLine(215,125)(310,125) 
\DashLine(215,25)(215,125){6}
\put(170,80){$3$}
\put(170,50){$1$}
\put(197,70){$2$}} \put(99,5){Triangle prototype 5}
\end{picture}
\end{center}
The possible internal particles (ordered from 1 to 3) are:
$$ \tilde{\chi}^-_a, \ H^0 (\phi^0) , \ \tilde{\chi}^+_b, 
     \ \ (a,b=1 \cdots 2) $$
$$ \tilde{\chi}^0_a, \ H^\pm (\phi^\pm) , \ \tilde{\chi}^0_b, 
     \ \ (a,b=1 \cdots 4) $$
$$ f, \ \tilde{f}, \ f $$
where $f$ stands for all matter fermions (quarks and leptons)
of either chirality
and $\tilde{f}$  their corresponding scalar super-partners.
\bigskip

\begin{center}
\begin{picture}(333,150)
\SetScale{0.8}
\footnotesize{
\Photon(0,75)(130,75){4}{12}
\ArrowLine(310,25)(215,25) 
\ArrowLine(215,25)(130,75)
\ArrowLine(130,75)(215,125)
\ArrowLine(215,125)(310,125) 
\Photon(215,25)(215,125){4}{10}
\put(170,80){$3$}
\put(168,50){$1$}
\put(192,70){$2$}}  \put(99,5){Triangle prototype 6}
\end{picture}
\end{center}
The possible internal particles (ordered from 1 to 3) are:
$$ \tilde{\chi}^0_a, \ W , \ \tilde{\chi}^0_b  ,  \ \ (a,b=1 \cdots 4) $$
$$ \tilde{\chi}^-_a, \ Z , \ \tilde{\chi}^+_b  ,  \ \ (a,b=1 \cdots 2) $$
\bigskip

\begin{center}
\begin{picture}(333,150)
\SetScale{0.8}
\footnotesize{
\Photon(0,75)(130,75){4}{12}
\ArrowLine(310,25)(215,25) 
\Photon(215,25)(130,75){4}{10}
\Photon(130,75)(215,125){4}{10}
\ArrowLine(215,125)(310,125) 
\ArrowLine(215,25)(215,125)
\put(170,82){$3$}
\put(170,53){$1$}
\put(195,70){$2$}}  \put(99,5){Triangle prototype 7}
\end{picture}
\end{center}
The internal particles (ordered from 1 to 3) are:
$$ W, \   \tilde{\chi}^0_a   \ W, \ \ (a=1 \cdots 4)    $$ 
\bigskip

\begin{center}
\begin{picture}(333,150)
\SetScale{0.8}
\footnotesize{
\Photon(0,75)(130,75){4}{12}
\ArrowLine(310,25)(215,25) 
\DashLine(215,25)(130,75){6}
\DashLine(130,75)(215,125){6}
\ArrowLine(215,125)(310,125) 
\ArrowLine(215,25)(215,125)
\put(170,80){$3$}
\put(170,52){$1$}
\put(190,70){$2$}}  \put(99,5){Triangle prototype 8}
\end{picture}
\end{center}
The possible internal particles (ordered from 1 to 3) are:
$$  \ H^0 (\phi^0) , \ \tilde{\chi}^+_a, \  H^0 (\phi^0)
     \ \ (a=1 \cdots 2) $$
$$  \ H^\pm (\phi^\pm) , \ \tilde{\chi}^0_a, \  H^\pm (\phi^\pm)
     \ \ (a=1 \cdots 4) $$
$$ \tilde{f}, \ f, \  \tilde{f}$$
where $f$ stands for all matter fermions (quarks and leptons)
of either chirality
and $\tilde{f}$  their corresponding scalar super-partners.
\bigskip

\begin{center}
\begin{picture}(333,150)
\SetScale{0.8}
\footnotesize{
\Photon(0,75)(130,75){4}{12}
\ArrowLine(310,25)(215,25) 
\DashLine(215,25)(130,75){6}
\Photon(130,75)(215,125){4}{10}
\ArrowLine(215,125)(310,125) 
\ArrowLine(215,25)(215,125)

\put(170,80){$3$}
\put(170,51){$1$}
\put(195,70){$2$}}  \put(99,5){Triangle prototype 9a}
\end{picture}
\end{center}
The possible internal particles (ordered from 1 to 3) are:
$$ H^0 (\phi^0), \  ,  \tilde{\chi}^-_a, \ Z,  \ \ (a=1 \cdots 2) $$
$$ H^\pm (\phi^\pm), \  ,  \tilde{\chi}^0_a, \ W,  \ \ (a=1 \cdots 4) $$
\bigskip

\begin{center}
\begin{picture}(333,150)
\SetScale{0.8}
\footnotesize{
\Photon(0,75)(130,75){4}{12}
\ArrowLine(310,25)(215,25) 
\Photon(215,25)(130,75){4}{10}
\DashLine(130,75)(215,125){6}
\ArrowLine(215,125)(310,125)
\ArrowLine(215,25)(215,125)
\put(170,80){$3$}
\put(170,53){$1$}
\put(195,70){$2$}}  \put(99,5){Triangle prototype 9b}
\end{picture}
\end{center}
The possible internal particles (ordered from 1 to 3) are:
$$ Z, \    \tilde{\chi}^-_a, \  H^0 (\phi^0),
  \ \ (a=1 \cdots 2) $$
$$ W, \    \tilde{\chi}^0_a, \  H^\pm (\phi^\pm),  \ \ (a=1 \cdots 4) $$
\bigskip

\subsection*{Box Diagrams}

\begin{center}
\begin{picture}(333,220)
\SetScale{0.8}
\footnotesize{
\ArrowLine(0,50)(100,50)\ArrowLine(100,50)(100,150) \ArrowLine(100,150)(0,150)
\ArrowLine(300,50)(200,50) \ArrowLine(200,50)(200,150)
\ArrowLine(200,150)(300,150) \Photon(100,50)(200,50){5}{7} 
 \Photon(100,150)(200,150){5}{7}
\put(145,60){$1$} \put(102,100){$2$}
\put(145,130){$3$} \put(204,100){$4$}}
\put(99,20){Box prototype 1}
\end{picture}

\end{center}
The possible internal particles (ordered from 1 to 4) are:
$$ Z, \ e, \ Z, \ \tilde{\chi}^-_a, \ \ (a=1 \cdots 2) $$ 
$$ W, \ \nu, \ W, \ \tilde{\chi}^0_a \ \ (a=1 \cdots 4) $$  
Each of these graphs is accompanied by a similar graph,
 crossed in the $s$-channel.
\bigskip

\begin{center}
\begin{picture}(333,200)
\SetScale{0.8}
\footnotesize{
\ArrowLine(0,50)(100,50)\ArrowLine(100,50)(100,150) \ArrowLine(100,150)(0,150)
\ArrowLine(300,50)(200,50) \ArrowLine(200,50)(200,150) 
\ArrowLine(200,150)(300,150) \DashLine(100,50)(200,50){5} 
 \DashLine(100,150)(200,150){5}
\put(145,60){$1$} \put(103,100){$2$}
\put(145,130){$3$} \put(203,100){$4$}}
\put(99,20){Box prototype 2}
\end{picture}
\end{center}
The possible internal particles (ordered from 1 to 4) are:
$$ \tilde{\nu}, \ \tilde{\chi}^-_a, 
   \ \tilde{\nu}, \  e, \ \ (a=1 \cdots 2) $$ 
$$ \tilde{e}, \ \tilde{\chi}^0_a, 
   \ \tilde{e}_L, \  \nu, \ \ (a=1 \cdots 4) $$ 
The first of these is accompanied by a similar graph,
 crossed in the $s$-channel.
\bigskip

\begin{center}
\begin{picture}(333,200)
\SetScale{0.8}
\footnotesize{
\ArrowLine(0,50)(100,50)\ArrowLine(100,50)(200,50) \ArrowLine(200,50)(300,50)
\ArrowLine(300,150)(200,150) \ArrowLine(200,150)(100,150)
 \ArrowLine(100,150)(0,150) 
 \DashLine(100,50)(100,150){5} 
 \Photon(200,150)(200,50){5}{5}
\put(145,60){$1$} \put(103,100){$2$}
\put(145,130){$3$} \put(204,100){$4$}} \put(99,20){Box prototype 3a}
\end{picture}
\end{center}
The possible internal particles (ordered from 1 to 4) are:
$$ \tilde{\chi}^0_a, \ \tilde{e}_j, \    W,
   \ \tilde{\chi}^0_b  \ \ (a,b=1 \cdots 4, \ \ j=L,R) $$ 
$$ \tilde{\chi}^-_a, \ \tilde{\nu}, \ W,
   \ \tilde{\chi}^-_b  \ \ (a,b=1 \cdots 2)  $$ 
The first of these is accompanied by a similar diagram crossed
in the $s$-channel.
\bigskip

\begin{center}
\begin{picture}(333,200)
\SetScale{0.8}
\footnotesize{
\ArrowLine(0,50)(100,50)\ArrowLine(100,50)(200,50) \ArrowLine(200,50)(300,50)
\ArrowLine(300,150)(200,150) \ArrowLine(200,150)(100,150)
 \ArrowLine(100,150)(0,150) 
\put(106,100){$2$}
 \Photon(100,50)(100,150){5}{5} 
 \DashLine(200,150)(200,50){5}
 
\put(145,60){$1$}
\put(145,130){$3$} \put(203,100){$4$}} \put(99,20){Box prototype 3b}
\end{picture}
\end{center}
The possible internal particles (ordered from 1 to 4) are:
$$ e,  \    W, \ e, 
   \ \tilde{\nu}  $$ 
\bigskip

\begin{center}
\begin{picture}(333,200)
\SetScale{0.8}
\footnotesize{
\ArrowLine(0,50)(100,50)\ArrowLine(100,50)(200,50) \ArrowLine(200,50)(300,50)
\ArrowLine(300,150)(200,150) \ArrowLine(200,150)(100,150)
 \ArrowLine(100,150)(0,150) 
\put(103,100){$2$}
 \DashLine(100,50)(100,150){5} 
 \DashLine(200,150)(200,50){5}
\put(145,60){$1$}
\put(145,130){$3$} \put(203,100){$4$}} \put(99,20){Box prototype 4}
\end{picture}
\end{center}
The possible internal particles (ordered from 1 to 4) are:
$$ \tilde{\chi}^0_a,  \    \tilde{e}_j, \  \tilde{\chi}^0_b, 
   \ H^\pm \, (\phi^\pm)    \ \ (a,b=1 \cdots 4, \ \ j=L,R) $$ 
 $$ \tilde{\chi}^-_a,  \    \tilde{\nu}, \  \tilde{\chi}^-_b, 
   \ H^0 \,(\phi^0)    \ \ (a,b=1 \cdots 2) $$ 
\bigskip

\begin{center}
\begin{picture}(333,200)
\SetScale{0.8}
\footnotesize{
\ArrowLine(0,50)(100,50)\ArrowLine(100,50)(200,50) \ArrowLine(200,50)(300,50)
\ArrowLine(300,150)(200,150) \ArrowLine(200,150)(100,150)
 \ArrowLine(100,150)(0,150) 
\put(113,85){$2$}
 \DashLine(100,50)(200,150){5} 
 \Photon(200,50)(100,150){4}{10}
\put(150,133){$3$} \put(174,85){$4$}}
\put(120,58){$\ \ \  \ \ 1$} \put(99,20){Box prototype 5a}
\end{picture}
\end{center}
The possible internal particles (ordered from 1 to 4) are:
$$ \tilde{\chi}^0_a,  \    \tilde{e}_L, \  \nu, \
   W    \ \ (a=1 \cdots 4) $$ 
$$ \tilde{\chi}^-_a,  \    \tilde{\nu}, \  e, \
   Z    \ \ (a=1 \cdots 2) $$ 
\bigskip

\begin{center}
\begin{picture}(333,200)
\SetScale{0.8}
\footnotesize{
\ArrowLine(0,50)(100,50)\ArrowLine(100,50)(200,50) \ArrowLine(200,50)(300,50)
\ArrowLine(300,150)(200,150) \ArrowLine(200,150)(100,150)
 \ArrowLine(100,150)(0,150) 
\put(111,85){$2$}
 \Photon(100,50)(200,150){4}{10} 
 \DashLine(200,50)(100,150){5}
\put(120,57){$\ \ \  \ \ 1$}
\put(150,133){$3$} \put(174,85){$4$}}
\put(120,58){$\ \ \  \ \ 1$} \put(99,20){Box prototype 5b}
\end{picture}
\end{center}
The possible internal particles (ordered from 1 to 4) are:
$$ \nu,  \    W, \  \tilde{\chi}^0_a, \
   \tilde{e}_L    \ \ (a=1 \cdots 4) $$ 
$$ e,  \    Z, \  \tilde{\chi}^-_a, \
   \tilde{\nu}    \ \ (a=1 \cdots 4) $$

\end{document}